\documentclass[11pt]{article}
\usepackage{amssymb}
\usepackage{amsmath}
\usepackage{graphics}
\usepackage{epsfig}
\usepackage{cite}
\usepackage{color}
\usepackage{mathrsfs}

\DeclareGraphicsExtensions{.eps}
\begin{document}

\title{ $ZZ\gamma$ production in the NLO QCD+EW accuracy at the LHC }
\author{ Wang Yong$^1$, Zhang Ren-You$^1$, Ma Wen-Gan$^1$, Li Xiao-Zhou$^1$, Wang Shao-Ming$^2$, and Bi Huan-Yu$^1$ \\
{\small $^1$ Department of Modern Physics, University of Science and Technology of China (USTC),}  \\
{\small  Hefei 230026, Anhui, People's Republic of China} \\
{\small $^2$ Department of Physics, Chongqing University, Chongqing 401331, People's Republic of China}
}

\date{}
\maketitle
\vskip 15mm
\begin{abstract}
In this paper we present the first study of the impact of the $\mathcal{O}(\alpha)$ EW correction to the $pp \to ZZ \gamma+X$ process at the CERN Large Hadron Collider (LHC). The subsequent $Z$-boson leptonic decays are considered at the leading order using the MadSpin method, which takes into account the spin-correlation and off-shell effects from the $Z$-boson decays. We provide numerical results of the integrated cross section and the kinematic distributions for this process. In coping with final-state photon-jet separation in the QCD real emission and photon-induced processes, we adopt both the Frixione isolated-photon plus jets algorithm and the phenomenological quark-to-photon fragmentation function method for comparison. We find that the next-to-leading order (NLO) EW correction to the $ZZ\gamma$ production can be sizeable and amounts to about $-7\%$ of the integrated cross section, and provides a non$-$negligible contribution to the kinematic distributions, particularly in the high energy region. We conclude that the NLO EW correction should be included in precision theoretical predictions in order to match future experimental accuracy.
\end{abstract}


\vfill \eject
\baselineskip=0.32in
\makeatletter      
\@addtoreset{equation}{section}
\makeatother       
\vskip 5mm
\renewcommand{\theequation}{\arabic{section}.\arabic{equation}}
\renewcommand{\thesection}{\Roman{section}}
\newcommand{\nb}{\nonumber}

\vskip 5mm
\section{Introduction}
\par
Probing the properties of Higgs boson, in particular its couplings to the standard model (SM) particles, is one of the significant missions of the current experiments at the CERN Large Hadron Collider (LHC), where the Higgs boson was discovered \cite{Higgs1,Higgs2}. Any possible deviation from the SM prediction means evidence of new physics. Therefore, the precise experimental measurements on Higgs boson properties are of top priority at the LHC Run2. Precise theoretical predictions are essential for both the signals and backgrounds of Higgs boson production to match experimental measurement accuracy. Triple gauge-boson productions at the LHC serves as a key process in studying quartic gauge-boson couplings (QGCs) and understanding electroweak (EW) symmetry spontaneous breaking. In the higher luminosity operation at the upgraded LHC, precision measurements become mandatory for decent predictions up to the QCD+EW next-to-leading order (NLO).

\par
$H \to Z\gamma$ decay is forbidden at tree level, but can be induced through a $W$-boson or massive quark loop in the SM. Thus an indication of new physics will be present if the particles circulating in the loop are not SM particles or the Higgs is a non-SM scalar boson, in particular at high-energy scale. The Higgs-radiation at the LHC, i.e., $pp \rightarrow ZH + X$, is a prominent channel to study the properties of the Higgs boson. After the sequential Higgs boson decay of $H \to Z\gamma$, the $ZZ\gamma$ production should be treated as an irreducible background of the $HZ$ production with subsequent $H \rightarrow Z \gamma$ decay and thus needs to be predicted with high precision. The publications \cite{ATLAS,CMS} provide a way to search for a Higgs boson decaying into $Z\gamma$ at the LHC, where a constraint was applied on the invariant mass of the $Z\gamma$ system around the Higgs boson mass to select the signal from this smooth production and other reducible backgrounds. In addition, the $ZZ\gamma$ production is also particularly interesting in investigating anomalous QGCs, such as $ZZ\gamma \gamma$ and $ZZZ\gamma$ couplings. Until now, the precision calculation for the $pp\to ZZ\gamma$ process at the LHC has been performed by including the NLO QCD correction \cite{ZZr-NLO-QCD}. A further step for the prediction combining both the NLO QCD and EW corrections is urgently requested, and is also the desired item in the 2013 and 2015 Les Houches high-precision wish lists \cite{wishlt2013,wishlt2015}.

\par
In this work, we report on a NLO QCD+EW correction to the $ZZ\gamma$ production at the $14~{\rm TeV}$ LHC, including the $Z$-boson leptonic decays with an improved narrow-width approximation(NWA). The rest of this paper is organized as follows: the calculation strategy is outlined in Section \ref{cal_detail}; numerical results of the integrated cross section and various kinematic distributions are presented in Section \ref{num_results}; finally, we give a short summary in Section \ref{summary}.

\vskip 5mm
\section{Calculation set-up }
\label{cal_detail}
\par
The NLO QCD and EW corrections to the parent process $pp\to ZZ\gamma+X$ are of $\mathcal{O}(\alpha_s \alpha^3)$ and $\mathcal{O}(\alpha^4)$, respectively. The NLO QCD correction has been presented in \cite{ZZr-NLO-QCD}, and in the following we mainly describe the calculation set-up for the NLO EW correction.

\par
\subsection{General set-up} \label{se:setup}
\par
\begin{figure*}[htbp]
\begin{center}
\includegraphics[scale=1.2]{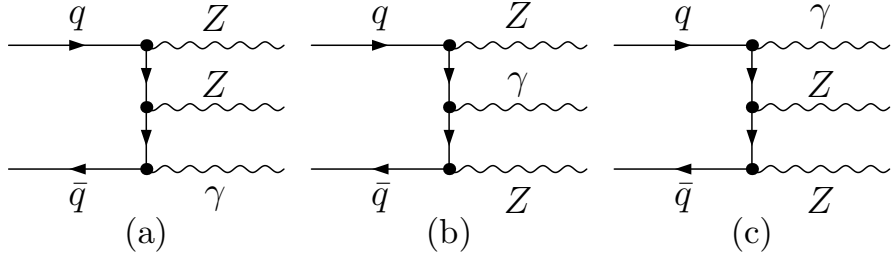}
\caption{\small The LO Feynman diagrams for the partonic process $q\bar{q}\to ZZ\gamma$.}
\label{fig1}
\end{center}
\end{figure*}

At the leading order (LO) the $ZZ\gamma$ events can only be produced via quark-antiquark annihilation at hadron colliders,
\begin{equation}
pp \rightarrow q\bar{q} \rightarrow ZZ\gamma + X~.
\label{born}
\end{equation}
Here we complete our calculation in the five-flavor scheme, i.e., $q=u,c,d,s,b$, and neglect all their masses.
The LO Feynman diagrams for the partonic process $q\bar{q} \rightarrow ZZ\gamma$ are shown in figure \ref{fig1}. The LO differential cross section for this partonic process is divergent in the phase-space region where the final-state photon is soft or collinear to one of the initial massless quarks. To avoid these LO infrared (IR) singularities and obtain an IR-safe result, we apply the following transverse momentum and rapidity cuts on the final photon:
\begin{eqnarray}
\label{constraints}
p_{T,\gamma}>20~{\rm GeV} ~~~~~{\rm and}~~~~~|y_\gamma|<2.5~.
\end{eqnarray}

\par
In the NLO calculation both the ultraviolet (UV) and IR singularities are regularized in dimensional regularization scheme, where the dimensions of spinor and space-time manifolds are extended to $D=4-2\epsilon$. We generate the related Feynman diagrams, simplify the amplitudes and perform numerical calculation successively by using the FeynArts-3.7+FormCalc-7.3+LoopTools-2.8 packages \cite{Feynarts,Formcalc, Looptools}. The IR singularities from the real jet and photon emissions are handled by both the Catani-Seymour dipole subtraction \cite{dipole} and the two cutoff phase-space slicing methods \cite{tcpss} for comparison. We use also the {\sc MadGraph5} \cite{Madgraph5} package to perform part of NLO QCD calculation, and find that the numerical results from both packages are coincident with each other within the calculation error.

\par
At the LO, the inputs of fine-structure constant are taken as $\alpha = \alpha(0)$ and $\alpha = \alpha_{G_{\mu}}$ for the electromagnetic and weak couplings, respectively. Then the LO squared amplitude is proportional to $\alpha(0)\alpha^2_{G_{\mu}}$. We calculate the NLO EW correction following the method in \cite{w+gamma,z+gamma}. In this way, the extra EW coupling at the EW NLO is chosen as $\alpha_{G_{\mu}}$, which is suitable for EW correction due to the EW Sudakov logarithms caused by the soft/collinear weak gauge-boson exchange at high energies \cite{Sudakov}.

\par
\subsection{EW virtual correction }
\par
The virtual correction is contributed by the related self-energy, vertex, box and pentagon Feynman diagrams. In the calculation of the NLO EW correction, the mixed scheme is used for EW couplings. As declared above, in this scheme the electromagnetic coupling in the LO amplitude is related to an $\alpha(0)$-scheme where $\alpha$ is defined in Thomson limit and the electric charge renormalization constant is thus \cite{alpha0-input}
\begin{eqnarray}
\label{detZe0}
\delta Z^{\alpha(0)}_e
= -\frac{1}{2}\delta Z_{AA} - \frac{1}{2} \tan\theta_W \delta Z_{ZA}=\left[
\frac{1}{2}\frac{\partial \sum^{AA}_T(p^2) }{\partial p^2} - \tan\theta_W \frac{\sum^{AZ}_T(p^2)}{M^2_Z}
\right]_{p^2=0}~,
\end{eqnarray}
where $\theta$ denotes the Weinberg angle and $\sum_T^{XY}(p^2)$ is the transverse part of the unrenormalized self-energy of $X\to{Y}$ transition at momentum squared $p^2$. As the mass-singular terms $\ln(m_f^2/\mu^2)~ (f = e,\mu,\tau,u,d,c,s,b)$ appear in both the electric charge renormalization constant and the external photonic wave-function counterterm, the exact cancelation between $\delta Z_e^{\alpha(0)}$ and $\frac{1}{2}\delta Z_{AA}$ helps to avoid these unpleasantly large logarithms. Finally, the remaining large logarithms can be absorbed into the running fine-structure constant in a $G_\mu$-scheme via
\begin{equation}
\alpha_{G_\mu} = \frac{\sqrt{2}G_{\mu}M_W^2(M_Z^2-M_W^2)}{\pi M_Z^2}~.
\end{equation}
The double counting in NLO EW correction is avoidable by modify the electric charge renormalization constant as
\begin{equation}\label{counterterm-Ze}
\delta Z_e^{G_\mu} = \delta Z_e^{\alpha(0)} -\frac{1}{2}\Delta r~,
\end{equation}
where $\Delta r$ is obtained from the one-loop EW correction to the muon decay \cite{delta}. In loop graphs shown in figure {\ref{fig2}, the divergent behavior in the vicinity of $M^2_{Z\gamma}=M^2_H$ can be handled by introducing the substitution $M_H^2 \to M_H^2-iM_H\Gamma_H$, while the contribution from the imaginary part is of higher order and small enough to be ignored. Actually, the interference term between the loop diagrams involving $H \rightarrow Z\gamma$ decay and the LO $q\bar{q} \rightarrow ZZ\gamma$ graphs is less than $0.2\%$.
\begin{figure*}[htbp]
\begin{center}
\includegraphics[scale=0.5]{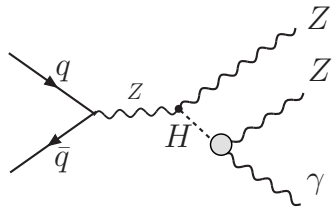}
\caption{\small Higgs resonance at the EW one-loop level. }
\label{fig2}
\end{center}
\end{figure*}

\par
\subsection{Real photon emission correction }
\par
\par
In the real photon emission at $\mathcal{O}({\alpha^4})$, two photons can be found in the final state through quark-antiquark scattering,
\begin{equation}
\label{real-gamma}
q\bar{q} \to ZZ\gamma\gamma~.
\end{equation}
Theoretically the IR divergences located at the phase-space region where one of the final-state photons tends to be soft or collinear to the incoming quark or antiquark. Those soft/collinear divergences are exactly/partially canceled by those from loop diagrams, and the remaining collinear IR singularities are absorbed by the EW counterterms of quark PDFs in the deep inelastic scattering (DIS) scheme as recommended in the EW NLO corrections using NNPDF2.3QED PDF set \cite{NNPDF}. If the two final-state photons are sufficiently collinear, they will be clustered into one quasi-photon.

\par
In this work, we dispose of all tensor and scalar integrals using the LoopTools-2.8 package \cite{Looptools}, in which the most complicated five-point integrals are decomposed into four-point integrals \cite{Denner} and then reduced to scalar integrals recursively by employing the Passarino-Veltman algorithm \cite{PV}. To solve the numerical instability induced by the small Gram determinant in the four-point integrals, we add a quadruple precision arithmetic switch in the LoopTools-2.8 package, adopting the similar method to that used in Refs.\cite{detG1,detG2}, which can automatically turn to the quadruple precision in case where the Gram determinant is sufficiently small.

\par
\subsection{Real gluon/light-quark emission and photon-induced corrections }
\par
At the QCD NLO the following generic tree-level processes contribute at $\mathcal{O}({\alpha^3\alpha_s})$:
\begin{align}
\label{real-g}
&q\bar{q} \to ZZ\gamma g~, \\
\label{real-q}
&qg \to ZZ\gamma q~.
\end{align}
The photon-induced process,
\begin{align}
\label{gamma-ind}
&q\gamma \to ZZ\gamma q~,
\end{align}
can also contributes at $\mathcal{O}({\alpha^4})$ due to the existence of the photon density inside the proton in the NNPDF2.3QED PDFs, and therefore should be taken into account. In equations (\ref{real-q}), (\ref{gamma-ind}) and the following expressions, the symbol ``$q$" can be quark $q$ or antiquark $\bar{q}$ with no ambiguity. These processes have the same signature (ZZ$\gamma$+jet) in the final state. The singularities arising from initial-state collinear splitting configurations with $q\to gq^*$ in process (\ref{real-g}) and $g/\gamma \to q\bar{q}^*$ in the gluon/photon-induced process (\ref{real-q})/(\ref{gamma-ind}) should be subtracted from the partonic cross sections and absorbed into the PDF redefinitions which can be realized with different factorization schemes. In using NNPDF2.3QED PDF set, we take the $\overline{\rm MS}$ factorization scheme with the counterterms for $q\to gq^*$ and $g \to q\bar{q}^*$ configurations in the NLO QCD corrections from (\ref{real-g}) and (\ref{real-q}), and DIS scheme for $\gamma \to q\bar{q}^*$ configuration in photon-induced correction from (\ref{gamma-ind}), respectively. One can find the corresponding quark counterterms with different factorization schemes for all stated collinear configurations in \cite{dipole,detG1,zzj}. As we know, both the NLO QCD correction to specific ZZ$\gamma$ production and the NLO EW correction to ZZ+jet production obtain contributions from the ZZ$\gamma$+jet production. Thus, a consistent photon-jet separation with respect of the IR singularity cancelation is essential.

\par
\subsection{Photon-jet separation and event selection criterion}
\label{photon-jet separation}
\par
The NLO EW and QCD real corrections originate from the emission of an additional external photon and QCD parton, respectively. In the circumstance that the extra particle is a jet, we should have an effective criterion to classify whether the final state is the $ZZ\gamma$ or $ZZ+\rm jet$ event. Normally, this may cause a bit confusion in an exclusive process.

\par
Phenomenologically, there are two production mechanisms for the final photon. One is the emission of a photon off a quark in the direct process, which is computable in perturbative quantum field theory. The other is that the photon is produced in the long-distance process which can be described by quark-to-photon fragmentation function $D_{q\to\gamma}(z_\gamma)$. In the collinear photon-jet system they are always recombined to a single quasi-particle. It is natural to identify the photon event by giving proper threshold on the hadronic energy in the collinear photon-jet system. Such a naive selection criterion, however, may destroy the cancelation of the IR singularities. In this paper we provide results with two different photon-jet separation methods in dealing with this situation.

{\bf a. Quark-to-photon fragmentation function}
\par
In this approach, the candidate photon and jet are treated equivalently. They are recombined into one photon-jet system if the two tracks of jet and photon are sufficiently collinear, i.e., $R_{\gamma\rm{j}} \le R_0$, where $R_{\gamma\rm{j}}= \sqrt{(y_{\rm{j}}-y_{\gamma})^2+(\phi_{\rm{j}}-\phi_{\gamma})^2}$ denotes the photon-jet separation in the pseudorapidity-azimuthal angle plane and $R_0$ is a pre-defined distance parameter (typically chosen around 0.5). Once the collinear photon-jet is recombined, it is treated as a $ZZ\gamma$ event if the energy fraction of photon inside photon-jet system exceeds a certain threshold, i.e, $z_\gamma=\frac{E_\gamma}{E_\gamma+E_{\rm{j}}}\ge z_\gamma^{\rm{cut}}$ (typically chosen to be larger than 0.9); otherwise it is rejected as $ZZ$+jet event. In this event identification criterion, some residual jet activities are always involved in the collinear photon-jet system and spoil the cancelation of the final-state collinear divergences for (\ref{real-q}) and (\ref{gamma-ind}) processes, since $z_\gamma$ ranges only from $z_\gamma^{\rm{cut}}$ to $1$. To obtain an IR-safe result, the photon fragmentation contribution should be taken into consideration with remaining QED collinear IR divergence absorbed into the NLO definition of the quark-to-photon fragmentation function, which is expressed in the $\overline{\rm{MS}}$ factorization scheme as \cite{LEP1}
\begin{eqnarray}
D^{{\rm bare}}_{q \rightarrow \gamma}(z_\gamma) =
\frac{Q_q^2 \alpha}{2\pi} \frac{1}{\epsilon} \frac{1}{\Gamma(1-\epsilon)}
\Big( \frac{4 \pi \mu_r^2}{\mu_f^2} \Big)^{\epsilon}
P_{\gamma q}(z_{\gamma}) + D_{q \rightarrow \gamma}(z_\gamma,\mu_f)~,
\end{eqnarray}
where $Q_q$ is the electric charge of the massless quark $q$ and the quark-to-photon splitting function $P_{\gamma q}(z_{\gamma})$ is given by
\begin{eqnarray}
P_{\gamma q}(z_{\gamma}) = \frac{1 + (1 - z_{\gamma})^2}{z_{\gamma}}~.
\end{eqnarray}
The nonperturbative quark-to-photon fragmentation function $D_{q \rightarrow \gamma}(z_\gamma,\mu_f)$ has been parametrized and the values of the fitting parameters have been fixed by the ALEPH collaboration at the LEP in analyzing $\gamma + n~{\rm jet}$ process \cite{LEP2}:
\begin{eqnarray}
D_{q \rightarrow \gamma}(z_\gamma,\mu_f) =
D_{q \rightarrow \gamma}^{{\rm ALEPH}}(z_\gamma,\mu_f) \equiv \frac{Q_q^2 \alpha}{2\pi}
\biggl( P_{\gamma q}(z_\gamma) \ln\frac{\mu^2_f}{(1-z_\gamma)^2 \mu^2_0} + C \biggr)~,
\end{eqnarray}
where $\mu_0 = 0.14~ {\rm GeV}$ and $C = -1 - \ln(M_Z^2/2\mu_0^2) = -13.26$ are obtained from one-parameter data fit. The collinear-safe effective quark-to-photon fragmentation function ${\cal D}_{q \rightarrow \gamma}(z_{\gamma})$ formulated in one-cutoff phase-space slicing method is established in \cite{LEP1}, written as
\begin{eqnarray}
\label{D-eff}
{\cal D}_{q \rightarrow \gamma}(z_\gamma) = -\frac{Q_q^2 \alpha}{2\pi} \frac{1}{\epsilon} \frac{1}{\Gamma(1-\epsilon)} \Big( \frac{4 \pi \mu_r^2}{\delta_c \hat{s}} \Big)^{\epsilon}
[ z_\gamma (1 - z_\gamma) ]^{-\epsilon} [ P_{\gamma q}(z_\gamma) - \epsilon z_{\gamma} ] +
D^{{\rm bare}}_{q \rightarrow \gamma}(z_\gamma)~.
\end{eqnarray}
The first term is the perturbatively calculable contribution in the region of $\hat{s}_{q\gamma}<\delta_c \hat{s}$, where $\hat{s}_{q\gamma}$ is the invariant mass of collinear quark-photon system, $\delta_c$ is an arbitrary small collinear cutoff and $\hat{s}$ is the partonic center-of-mass colliding energy squared. Thus the fragmentation contribution to the partonic process (\ref{real-q}) or (\ref{gamma-ind}) is given by
\begin{eqnarray}
\label{frag}
d\hat{\sigma}_{\rm{frag}}(qg[\gamma]\to ZZ \gamma q)=d\hat{\sigma}^{\rm{LO}}(qg[\gamma]\to ZZq)\int^1_{z_\gamma^{{\rm{cut}}}} dz_\gamma {\cal D}_{q \rightarrow \gamma}(z_\gamma)~.
\end{eqnarray}

{\bf b. Frixione isolation method }
\par
In the Frixione isolation method \cite{Frixione} the threshold of jet energy in the collinear photon-jet system varies along with their distance $R_{\gamma \rm{j}}$ where only a soft parton can become collinear to the photon. The event is accepted only if
\begin{eqnarray}
\label{Frixione}
p_{T,\rm j} \le \chi(R_{\gamma {\rm j}}) ~~~~~{\rm or}~~~~~ R_{\gamma {\rm j}} > R_0=0.5~,
\end{eqnarray}
where $\chi(R_{\gamma {\rm j}})$ is an appropriate restriction function described by isolation parameter $\epsilon_\gamma$ and weight factor $n$ \cite{Frixione}. In this paper we set $\epsilon_\gamma=1$ and $n=1$. In the case of $R_{\gamma {\rm j}}\to 0$,  $\chi(R_{\gamma {\rm j}})$ is required to be going to zero as
\begin{equation}\label{Parameter-in-Frixione}
\chi(R_{\gamma {\rm j}}) = \Big(\frac{1-\cos R_{\gamma {\rm j}}}{1-\cos R_0}\Big)^n p_{T,\gamma}\epsilon_{\gamma}~.
\end{equation}
By this subtle construction, the Frixione isolation criterion forbids any hard collinear jet activity and therefore photon fragmentation contribution is eliminated. Meanwhile, the soft jet activity is retained to guarantee the cancelation of IR singularities between virtual and gluon emission corrections in the NLO QCD calculation of $ZZ\gamma$ production.

\par
For the $ZZ\gamma\gamma$ final state, the two photon tracks will be recombined into a quasi-particle if they are not well separated, i.e.,
\begin{eqnarray}
R_{\gamma\gamma}<0.1~,
\end{eqnarray}
and the event is regarded as $ZZ\gamma$. The $ZZ\gamma+{\rm jet}$ final state is also treated as a $ZZ\gamma$ event if $R_{\gamma {\rm j}}<R_0 = 0.5$ and $z_\gamma \ge z_\gamma^{{\rm{cut}}}=0.9$ in the quark-to-photon fragmentation function method. The kinematic constraints of (\ref{constraints}) are applied on these $ZZ\gamma$ events after performing the recombination procedure. In addition to the inclusive results for the jet activity, we also present exclusive results with a hard jet veto condition of $p_{T,\rm j}>p_{T,\rm j}^{\rm cut}=100$ GeV in this work as well.

\vskip 5mm
\section{Numerical results and discussion}
\label{num_results}
\par
\subsection{Input parameters}
We obtain all numerical results with the following SM input parameters from \cite{PDG}:
\begin{eqnarray}
&&
M_W = 80.385~ {\rm GeV}~,~~~~~
M_Z = 91.1876~ {\rm GeV}~,~~~~~
m_t = 173.21~ {\rm GeV}~,\nonumber \\
&&
M_H=125.09~{\rm GeV}~,~~~~
\Gamma_H=4~{\rm MeV}~,~~~~
G_{\mu} = 1.1663787 \times 10^{-5}~ {\rm GeV}^{-2}~,\nonumber \\
&&
\alpha(0)=1/137.035999139~.
\end{eqnarray}
For the $ZZ\gamma$ production at the LHC, the Cabibbo-Kobayashi-Maskawa (CKM) matrix elements are only involved in the EW one-loop amplitude. Since all the light quarks are treated as massless particles and the five-flavor scheme is adopted in the PDF convolution, the CKM matrix can be set to identity in our calculation if it is a $2 \oplus 1$ block-diagonal matrix. We use the NNPDF2.3QED PDF set in both the LO and NLO calculations. As discussed in \cite{NNPDF,DIS-nsct}, the NNPDF2.3QED PDF set is only of LO with respect to QED correction and the absence of EW correction in the PDF fit to data favors the DIS factorization scheme in the NLO EW calculation. Therefore, we employ the $\overline{\rm MS}$ and DIS schemes in the NLO QCD and NLO EW (also photon-induced process) calculations, respectively. The NLO QCD correction to $pp \to ZZ\gamma+X$ can be sketched as
\begin{eqnarray}
\big[\Delta\sigma^{\rm NLO}_{\rm QCD}\big]_{pp}^{\rm Frix(frag)}=
\big[\big( \sigma^0 - \sigma^{\textrm{LO}} \big) +
\sigma_{\rm{virt}}^{\alpha_s} +
\sigma_{\textrm{real}}^{\alpha_s} +
\sigma_{\textrm{pdf}}^{\alpha_s}\big]_{pp}^{\rm Frix(frag)} +
\big(\sigma_{\rm frag}^{\alpha_s}\big)_{qg}^{(\rm frag)}~.
\label{corrected_QCD}
\end{eqnarray}
The superscripts ``Frix" and ``frag" in equation (\ref{corrected_QCD}) stand for the two different photon-jet separation methods discussed in section \ref{photon-jet separation}, i.e., the abbreviations of Frixione isolation and quark-to-photon fragmentation function methods respectively. In the Frixione isolation method, the summation of all individual divergent pieces of $\sigma_{\textrm{virt}}^{\alpha_s}$, $\sigma_{\textrm{real}}^{\alpha_s}$ and $\sigma_{\textrm{pdf}}^{\alpha_s}$ originating from the virtual, real and PDF-counterterms at ${\cal O}(\alpha^3 \alpha_s)$ is proved to be finite. Once the quark-to-photon fragmentation method is applied, the photon fragmentation contribution in the gluon-induced subprocess (\ref{real-q}), which corresponds to the last piece in parenthesis in (\ref{corrected_QCD}), should be included to make an IR-safe prediction. $\sigma^{\textrm{LO}}$ and $\sigma^{0}$ are the LO cross sections calculated by using the LO and NLO NNPDF2.3QED PDF separately. Similarly, the NLO EW and photon-induced corrections can be decomposed as
\begin{eqnarray}
&&
\big[\Delta\sigma^{\rm NLO}_{\rm EW}\big]_{q\bar{q}}=
\big[\sigma_{\rm{virt}}^{\alpha} +
\sigma_{\textrm{real}}^{\alpha} +
\sigma_{\textrm{pdf}}^{\alpha}\big]_{q\bar{q}}~, \nonumber \\
&&
\big[\Delta\sigma^{\rm NLO}_{\gamma\textrm{-ind}}\big]_{q\gamma}^{\rm Frix(frag)}=
\big[\sigma_{\textrm{real}}^{\alpha} +
\sigma_{\textrm{pdf}}^{\alpha}\big]_{q\gamma}^{\rm Frix(frag)} +
\big(\sigma_{\rm frag}^{\alpha}\big)_{q\gamma}^{(\rm frag)}~.
\end{eqnarray}
Unlike the NLO QCD correction, the NLO EW correction via quark-antiquark annihilation (\ref{real-gamma})  avoids the photon-jet separation problem. While the photon-induced channel (\ref{gamma-ind}) like the gluon-induced channel (\ref{real-q}) needs the final-state photon-jet separation methods. We define the relative QCD, EW and photon-induced corrections as follows:
\begin{eqnarray}
\delta_{\rm QCD}=\frac{\Delta\sigma_{\rm QCD}^{\rm NLO}}{\sigma^{\rm LO}}~,~~~~~~
\delta_{\rm EW}=\frac{\Delta\sigma_{\rm EW}^{\rm NLO}}{\sigma^{0}}~,~~~~~~
\delta_{\gamma\textrm{-ind}}=\frac{\Delta\sigma_{\gamma\textrm{-ind}}^{\rm NLO}}{\sigma^{0}}~.
\end{eqnarray}
There the relative NLO EW and photon-induced corrections are normalized by $\sigma_0$ in order to eliminate the QCD contributions from the NLO PDFs. The NLO QCD and EW corrected cross section can be obtained by combining the LO cross section, the QCD, EW and photon-induced corrections together, either by using additive approximation (denoted as QCD$\oplus$EW),
\begin{eqnarray}
\sigma_{\rm QCD\oplus EW}^{\rm NLO} &=& \sigma^{\rm LO}(1+\delta_{\rm QCD\oplus EW}) \nonumber \\
&=& \sigma^{\rm LO}(1 + \delta_{\textrm{QCD}} + \delta_{\textrm{EW}} + \delta_{\gamma\textrm{-ind}})~,
\label{add-approach}
\end{eqnarray}
or multiplying approximation (denoted as QCD$\otimes$EW) to identify the high order interplay of QCD and EW corrections,
\begin{eqnarray}
\sigma_{\rm QCD\otimes EW}^{\rm NLO} &=& \sigma^{\rm LO}(1+\delta_{\rm QCD\otimes EW}) \nonumber \\
&=& \sigma^{\rm LO}\Big[( 1 + \delta_{\textrm{QCD}} ) ( 1+ \delta_{\textrm{EW}}) + \delta_{\gamma\textrm{-ind}} \Big]~ ,
\label{multi-approach}
\end{eqnarray}
which can be regard as an improved prediction in case that the QCD and EW scales are clearly separated. We apply mainly the multiplying approximation in follow calculations, except in discussing the additional uncertainty due to adopt these two approaches. This uncertainty in the combined QCD and EW relative corrections can be roughly estimated by the difference of  the results from above two approximations. In studying the scale dependence of the NLO QCD predictions, we set the factorization and renormalization scales to be equal throughout our calculations for simplicity, and choose the central scales as
\begin{eqnarray}
\mu_0 = H_T/2 = \sum\limits_{i} m_{T,i}/2~,
\end{eqnarray}
where $m_{T,i} = \sqrt{m_i^2 + \vec{p}_{T,i}^{\,2}}$ is the transverse mass of the final-state particle $i$ and the summation is taken over all the final particles for the process $pp \to ZZ\gamma + X$. In the following numerical results, we take $\mu_r =\mu_f =\mu_0$ by default unless otherwise stated.

\subsection{Integrated cross sections}
\par
To show the residual dependence of the integrated cross section on scale parameters originating from fixed NLO truncation calculations, we study the scale uncertainty related to factorization and renormalization scale by setting $\mu_f=\mu_r=\mu$ with $\mu$ varying around the central scale $\mu_0$ as $\mu=x\mu_0$, $x \in [0.5,~2]$. The graphic presentation over scale dependence is shown in figure \ref{fig3}, where we observe that the LO scale dependence is slightly reduced by including the NLO QCD$\otimes$EW corrections either in Frixione isolation or quark-to-photon fragmentation function scheme. When the hard jet-veto condition $p_{T,{\rm j}}>p_{T,\rm{j}}^{\rm{cut}}=100~\rm{GeV}$ is applied, the integrated cross section is then considerably diminished and the scale dependence is also a little bit weakened. Obviously, this is because in jet-veto scheme we abort a lot of events which are from the real emission processes.
However, the jet veto would induce an additional theoretical uncertainty due to the logarithmic terms of $\ln(p^{\, {\rm cut} \, \, 2}_{T,\, {\rm jet}}/\mu^2)$ in the exclusive selection scheme. This theoretical uncertainty can be  principally improved by the resummation of these large logarithms \cite{jt-uncertainty}, which is beyond the scope of our paper.
\begin{figure*}[htbp]
\begin{center}
\includegraphics[width=12cm]{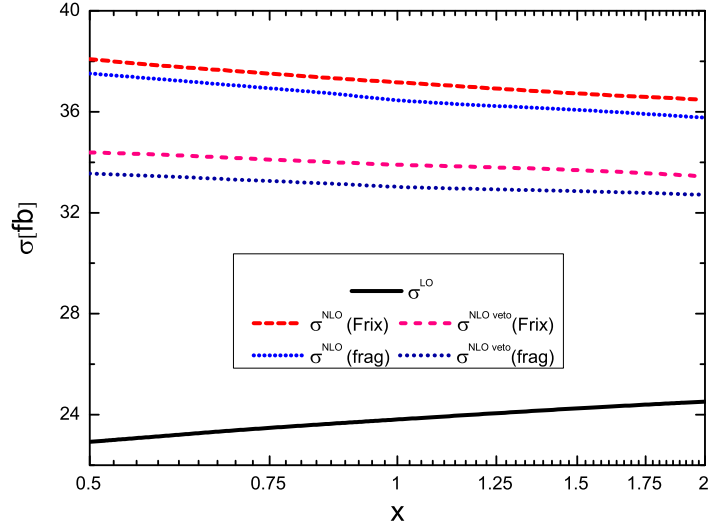}
\caption{\small The scale dependence of the LO and NLO QCD$\otimes$EW integrated cross sections for the $pp\to ZZ\gamma+X$ production at the 14 TeV LHC by using two photon-jet isolation prescriptions (see section \ref{photon-jet separation}) with and without jet veto.}
\label{fig3}
\end{center}
\end{figure*}

\begin{table}[htbp]
\begin{center}
\renewcommand\arraystretch{1.5}
\begin{tabular}{|c|c|c|c|c|}
\hline   $\rm{\sqrt{s}=14~TeV}$      &\multicolumn{2}{|c|}{Frixione isolation}      &\multicolumn{2}{|c|}{fragmentation function} \\
\cline{2-5}                     &$\rm{inclusive}$     &$\rm{exclusive}$                  &$\rm{inclusive}$ &$\rm{exclusive}$  \\
\hline
$\sigma^{\rm LO}~[fb]$             &\multicolumn{4}{|c|}{$23.816(3)_{-3.7\%}^{+3.0\%}$} \\
\hline
$\delta_{\rm EW}~[\%]$             &\multicolumn{4}{|c|}{-6.8}  \\
\hline
$\delta_{\gamma{\textrm{-ind}}}~[\%]$  &0.02    &0.004    &0.02    &0.002\\
\hline
$\delta_{\rm QCD}~[\%]$            &67.4    &52.6    &63.9    &48.6 \\
\hline
$\delta^{\rm NLO}_{\rm QCD \otimes \rm EW}~[\%]$            &56.1    &42.2    &52.8    &38.5 \\
\hline
$\sigma^{NLO}_{\rm QCD \otimes \rm EW}~[fb]$       &$37.18(4)_{-1.9\%}^{+2.4\%}$   &$33.87(4)_{-1.3\%}^{+1.5\%}$   &$36.39(2)_{-1.8\%}^{+3.1\%}$   &$32.99(2)_{-1.4\%}^{+1.2\%}$ \\
\hline
\end{tabular}
\caption{\small  \label{tab3}
The LO, NLO QCD$\otimes$EW integrated cross sections and the corresponding NLO EW, NLO QCD, NLO QCD$\otimes$EW and photon-induced relative corrections at the 14 TeV LHC by taking $p_{T,\gamma}>p_{T,\gamma}^{\rm{cut}}=20~{\rm GeV}, |y_\gamma|<2.5$. The results are given in two photon-jet separation schemes (see section \ref{photon-jet separation}) by applying the jet-veto condition $p_{T,\rm{j}}>p_{T,\rm{j}}^{\rm{cut}}=100~{\rm GeV}$ (exclusive scheme) as well as without it (inclusive scheme). For $\sigma^{\rm LO}$ and $\sigma^{\rm NLO}_{\rm QCD \otimes \rm EW}$, the central values represent the total cross sections obtained by taking $\mu=\mu_0$, and the upper and lower relative errors are the maximal and minimal relative uncertainties with $\mu$ varying in the range of $0.5\mu_0 \le \mu \le 2 \mu_0$.}
\label{tab1}
\end{center}
\end{table}
\par
In table \ref{tab1} we give the LO, NLO QCD$\otimes$EW corrected cross sections and corresponding NLO QCD, EW, QCD$\otimes$EW and photon-induced relative corrections at the $\sqrt{s}=14$ TeV LHC. We present numerical results for the QCD real emission and photon-induced corrections by taking Frixione isolation/fragmentation function photon-jet separation criterion and inclusive/exclusive event selection schemes separately. We see that for the integrated cross sections, the NLO QCD correction is enhanced, while the EW correction suppresses the LO cross section. The NLO EW correction is one order smaller, but quantitatively non$-$negligible. Moreover the significant EW correction always shows up in the kinematic distributions due to the EW Sudakov effect, which we will show in section \ref{kin distribution}. We can see from the table that the difference between the QCD relative corrections by using the two photon-jet separation schemes is about $5\%\sim 8\%$. When the events are discarded under the jet-veto condition $p_{T,\rm{j}}>p_{T,\rm{j}}^{\rm{cut}}=100$ GeV, the QCD correction is dropped by roughly $15\%$. Recently a new parametrization of the photon PDF, $\rm{LUXqed}\_\rm{PDF4LHC15}\_\rm{nnlo}\_100$ PDF set, is available \cite{LUXqed}. For comparison we also use that photon PDF set to calculate the photon-induced contributions and find that the results are even smaller than the corresponding ones by using the NNPDF2.3QED set. We see that the photon-induced contributions using both two PDF sets are phenomenologically negligible compared to all the other NLO correction components. The difference of $\delta^{\rm NLO}_{\rm QCD\oplus \rm EW}$ and $\delta^{\rm NLO}_{\rm QCD\otimes \rm EW}$ from equations (\ref{add-approach}) and (\ref{multi-approach}) can be used to estimate the uncertainty due to neglecting the higher order interplay between NLO QCD and NLO EW corrections. The NLO QCD+EW predictions in using additive and multiplying approachs can be simply estimated from table \ref{tab1} which differ by 2.5\% to 3\%. In the table the values of $\sigma^{\rm LO}$ and $\sigma^{\rm NLO}_{\rm QCD \otimes \rm EW}$ are the total cross sections obtained by taking $\mu=\mu_0$, and the upper and lower relative scale uncertainties ($\eta_{+}$ and $\eta_{-}$) are defined as
\begin{eqnarray}
\eta_{+(-)}=
\textrm{max(min)} \left( \frac{\sigma(\mu)}{\sigma(\mu_0)}-1\right) \, \biggl|_{\mu \in  [0.5\mu_0,~ 2 \mu_0 ]}~.
\end{eqnarray}

\subsection{Kinematic distributions} \label{kin distribution}
\par
In this subsection, we present some LO, NLO QCD and QCD+EW corrected kinematic distributions of final particles before and after the final $Z$-boson leptonic decays at the $14~ {\rm TeV}$ LHC. Since there is no significant distribution difference between the line-shapes when using the Frixione photon-jet separation algorithm and the quark-to-photon fragmentation function method, we only provide the distributions in the Frixione isolation scheme in the following discussion.

\subsubsection{ Kinematic distributions of Z-bosons }
\par
In figures \ref{fig4}(a), (b) we depict the LO, NLO QCD and NLO QCD+EW corrected invariant mass distributions of a $Z$-boson pair with and without jet veto ($p_{T,\rm{j}}^{\rm{cut}}=100$ GeV) (i.e., in the exclusive and inclusive event selection schemes). The corresponding relative corrections as well as scale uncertainties of NLO QCD$\otimes$EW relative corrections are shown in the lower panels (The EW correction introduces no renormalization scale dependence and its factorization scale dependence is much smaller than QCD one. Hence the scale dependence is mainly referred to as the QCD scale dependence.) The numerical NLO QCD+EW combined relative corrections are obtained by using both the additive and multiplying approximations to show the theoretical uncertainty due to missing higher order combined QCD and EW contributions. In these figures, all the LO, NLO QCD and combined NLO QCD+EW corrected invariant mass distributions reach their peaks in the vicinity of $M_{ZZ}\sim 200$ GeV and then descend with the increment of $M_{ZZ}$. The LO distribution is enhanced by the NLO QCD correction but suppressed by the NLO EW correction. Quantitatively the NLO QCD+EW combined correction is mainly from the NLO QCD correction, while the NLO EW correction provides less a contribution. However, in the virtue of the so-called EW Sudakov logarithm, the negative EW contribution in high $M_{ZZ}$ region becomes to be more meaningful. (e.g., we obtain $\delta_{\rm EW}\approx -10\%$ at $M_{ZZ}=450$ GeV). We can read from the figures that the inclusive relative NLO QCD correction increases slowly from $\sim 60\%$ to $\sim 75\%$ in the plotted invariant mass region, and the NLO QCD$\oplus$EW and NLO QCD$\otimes$EW combined relative corrections are roughly steady at about $60\%$ and $55\%$, respectively. With part of the QCD correction removed by taking jet-veto condition, the exclusive QCD relative correction is stable at about $50\%$, and the QCD$\otimes$EW combined relative correction is decreasing, varying from $\sim 50\%$ at threshold to $\sim 40\%$ at $M_{ZZ}=450$ GeV. The missing higher order combined QCD and EW corrections, which are NNLO contributions, can be estimated by the difference between two extreme combinations of the NLO QCD and EW corrections. The disparities between adopting two combination approximations can be differ by a few percent (about $5\%$ see Figure 4(a), (b)) due to the large QCD corrections and EW Sudakov logarithm in the invariant mass tail. We see that the difference is comparable with corresponding QCD scale uncertainty, thus both theoretical uncertainties should be considered equally.
\begin{figure}[htbp]
\includegraphics[width=9cm]{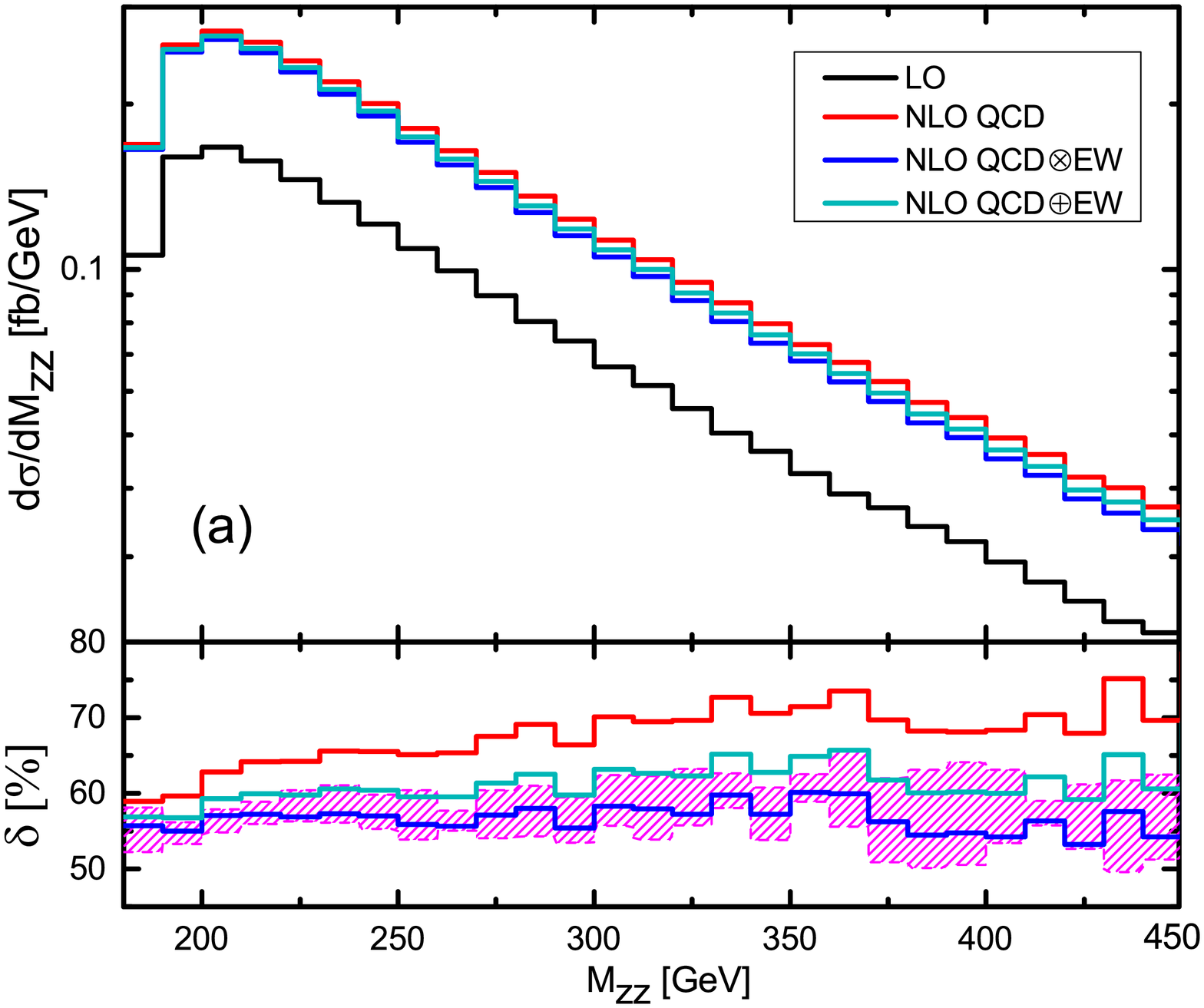}
\hspace{-1cm}
\includegraphics[width=9cm]{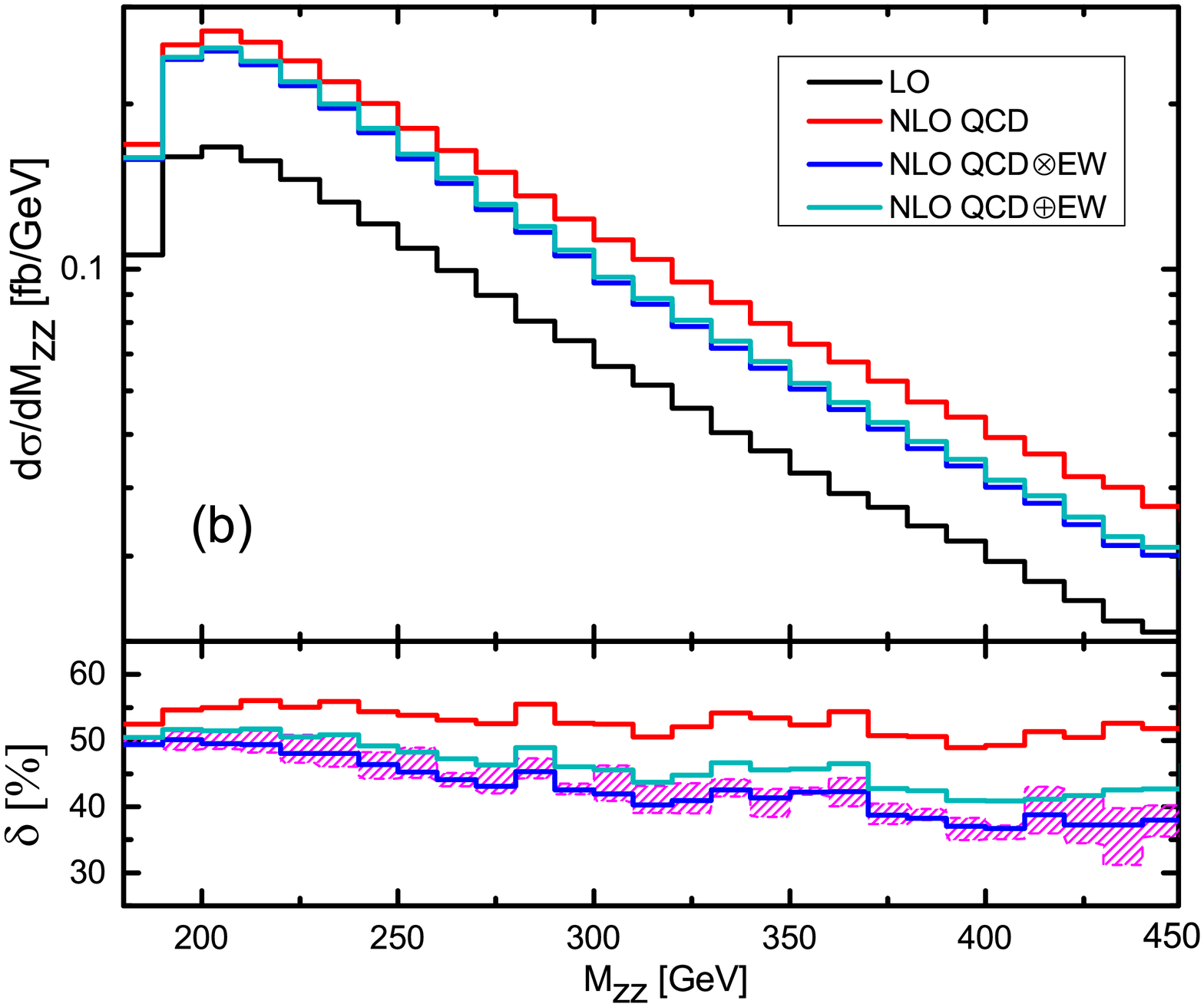}
\caption{ The $Z$-boson pair invariant mass distributions at the LO, including NLO QCD and NLO QCD+EW corrections for the inclusive (a) and exclusive (b) $ZZ\gamma$ production at the 14 TeV LHC. In the lower panels the pink bands are for the scale uncertainty ranges of the NLO QCD$\otimes$EW relative corrections, and the red, blue and green lines are for the NLO QCD, NLO QCD$\otimes$EW and NLO QCD$\oplus$EW relative correction distributions, respectively.}
\label{fig4}
\end{figure}

\par
We specify the final two $Z$-bosons as the leading $Z$-boson denoted as $Z_1$ if $p_{T,Z_1} > p_{T,Z_2}$, and the next-to-leading $Z$-boson $Z_2$ for another. In figures \ref{fig5}(a), (b) and figures \ref{fig6}(a), (b) we present the LO, NLO QCD, NLO QCD$\otimes$EW and NLO QCD$\oplus$EW corrected transverse momentum distributions of $Z_1$- and $Z_2$-boson separately. In the lower panels the corresponding relative correction distributions of final $Z$-bosons are depicted, and the scale uncertainty ranges for the NLO QCD$\otimes$EW relative corrections are plotted also in pink bands. Sharing the similarity with the $Z$-boson pair invariant mass distributions in figures \ref{fig4}(a), (b), the obvious negative EW corrections in high $p_T$ range can be seen for both the leading and next-to-leading $Z$-boson $p_T$ distributions originating from Sudakov-type high-energy logarithmic terms. In the exclusive scheme, a lot of events with energetic final jet are abandoned, that decreases the collected events number with high-energy $Z$-boson directly. We see that in figures \ref{fig5}(a), (b) all the LO and NLO corrected distributions are peaked at $p_{T,Z_1}\sim 50$ GeV for the leading $Z$-boson, and in figures \ref{fig6}(a), (b) the peaks are located at $p_{T,Z_2}\sim 20$ GeV for the next-to-leading $Z$-boson. Figure \ref{fig5}(a) shows that in the inclusive case for the $Z_1$-boson, the QCD relative correction increases from about $60\%$ to $110\%$ when $p_{T,Z_1}$ goes up in the range of $[50,300]$ GeV, while the NLO EW relative correction can reach about $-20\%$ in high $p_{T,Z_1}$ region, which makes the NLO QCD$\otimes$EW relative correction to be about $70\%$ at $p_{T,Z_1}=300$ GeV. We can see from figure \ref{fig5}(b) that in case of exclusive scheme, the jet-veto condition reduces the correction even more strongly than the EW correction part in large $p_{T,Z_1}$ region, while at the position of $p_{T,Z_1}=300$ GeV the exclusive QCD relative correction is nearly $30\%$. Figure \ref{fig5}(a) shows the NLO QCD$\otimes$EW and NLO QCD$\oplus$EW relative corrections in the inclusive scheme have a sizeable disparity in the tail of $p_{T,Z_1}$ (reaching about $20\%$). In the exclusive case (see figure \ref{fig5}(b)), the NLO QCD$\otimes$EW and NLO QCD$\oplus$EW combined relative corrections in the tail become fairly consistent owing to the fewer events collected due to the jet-veto constraint directly. However, these jet-veto discarded events are irreverent with the EW correction.

\par
Similarly, we can see from figures \ref{fig6}(a), (b) that the NLO QCD relative correction distribution of $Z_2$-boson receives sizeable reduction in high $p_{T,Z_2}$ region due to the effect of the EW Sudakov logarithms (which can reach about $-20\%$) for both the inclusive and exclusive cases. The QCD relative correction distribution for the $p_{T,Z_2}$ in the inclusive scheme remains stable at about $65\%$, while the inclusive QCD$\otimes$EW relative correction has a downtrend from $60\%$ to $30\%$. In the exclusive scheme the NLO QCD relative correction falls from $60\%$ to $30\%$ and the NLO QCD$\otimes$EW relative correction decreases gradually from $\sim 50\%$ to $\sim 5\%$ in the plotted $p_{T,Z_2}$ region. The difference between the heuristic NLO QCD$\otimes$EW and NLO QCD$\oplus$EW approximations can also be observed in $p_{T,Z_2}$ distribution, and is quantitatively comparable with the corresponding QCD scale uncertainty.

\begin{figure}[htbp]
\includegraphics[width=9cm]{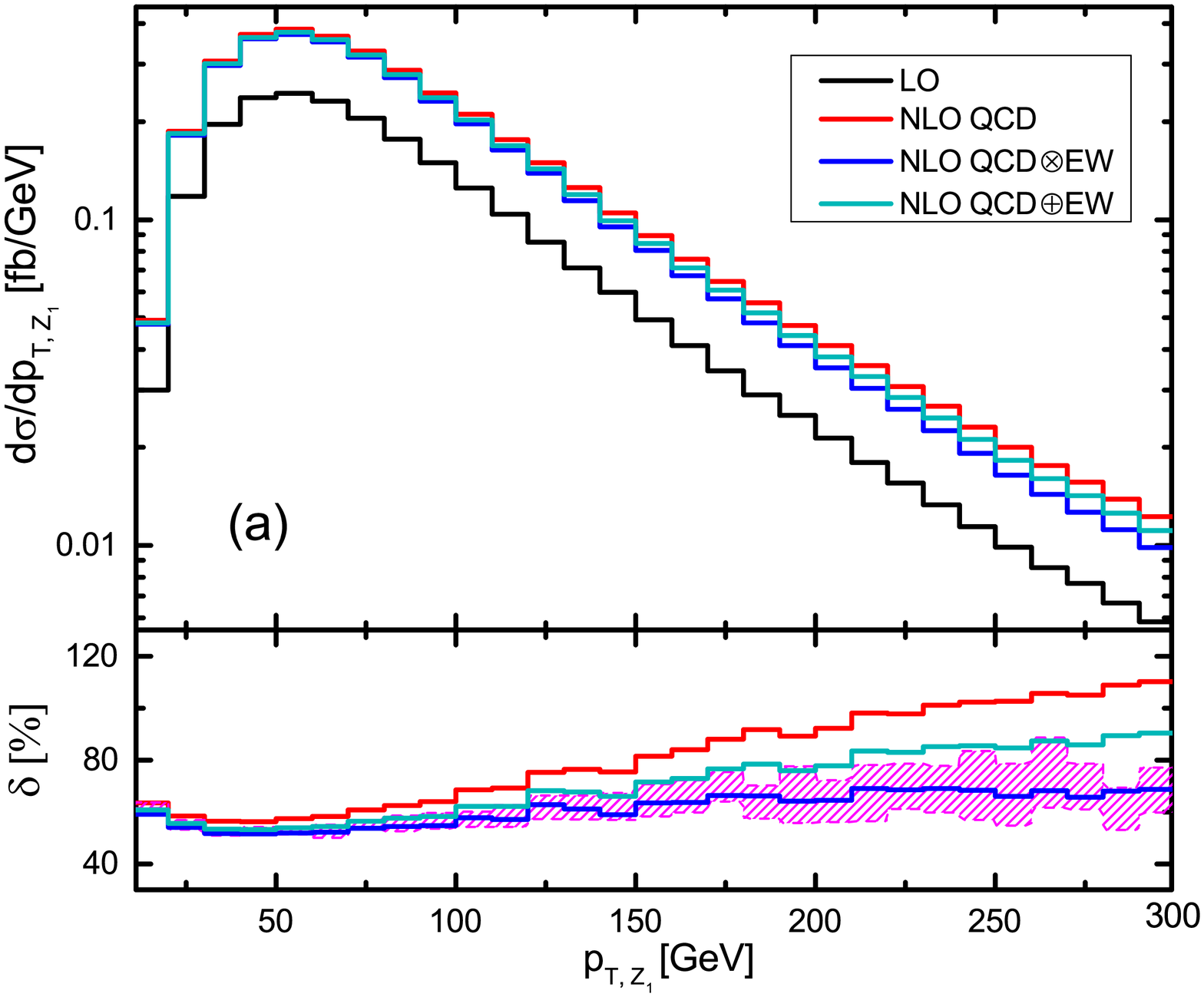}
\hspace{-1cm}
\includegraphics[width=9cm]{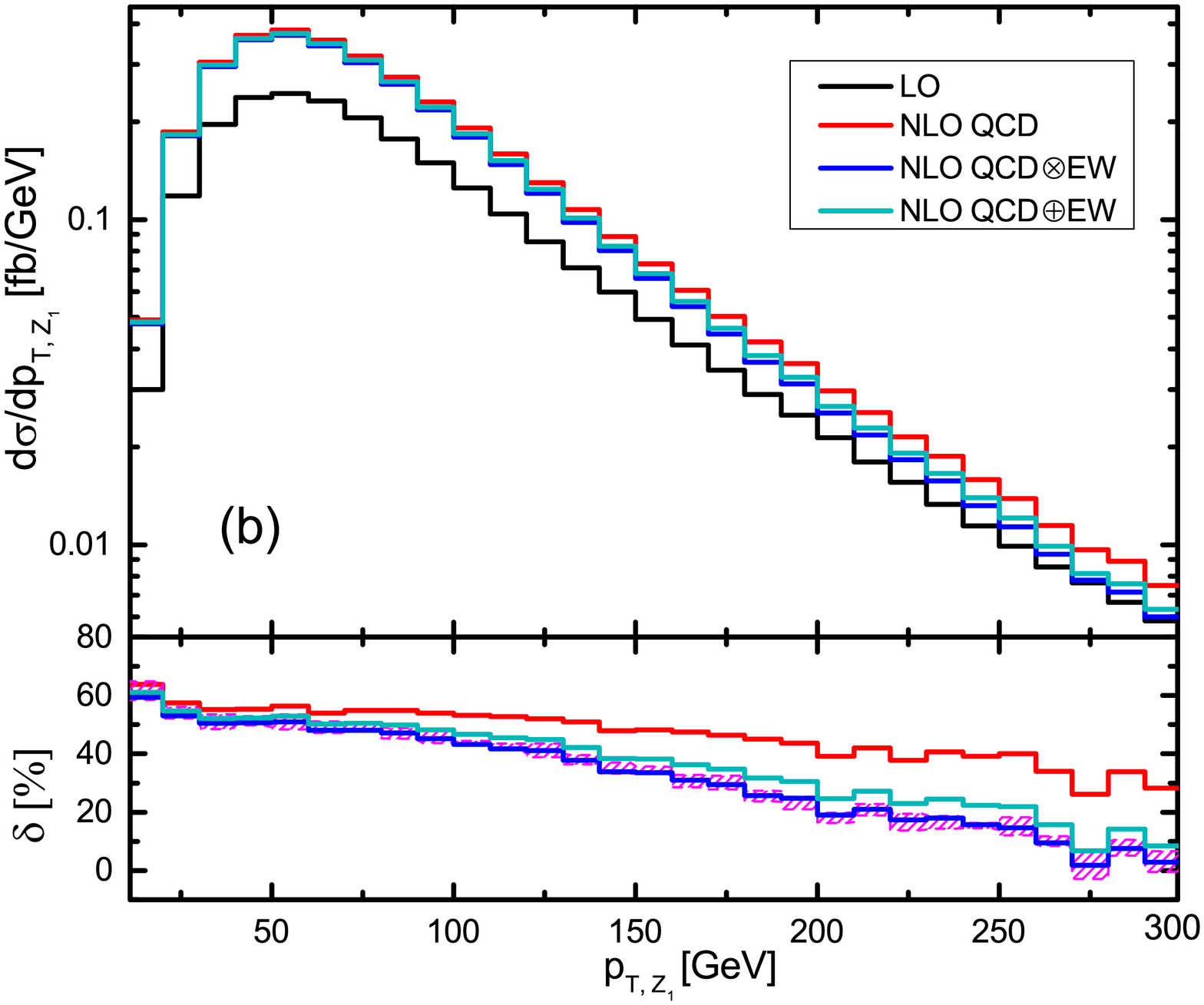}
\caption{ The LO, NLO QCD and NLO QCD+EW corrected distributions of the leading $Z$-boson transverse momentum distributions for the inclusive (a) and exclusive (b) $ZZ\gamma$ production at the 14 TeV LHC. In the lower panels the pink bands are for the scale uncertainty ranges of the NLO QCD$\otimes$EW relative corrections, and the red, blue and green lines are for the NLO QCD, NLO QCD$\otimes$EW and NLO QCD$\oplus$EW relative correction distributions, respectively. }
\label{fig5}
\end{figure}
\begin{figure}[htbp]
\includegraphics[width=9cm]{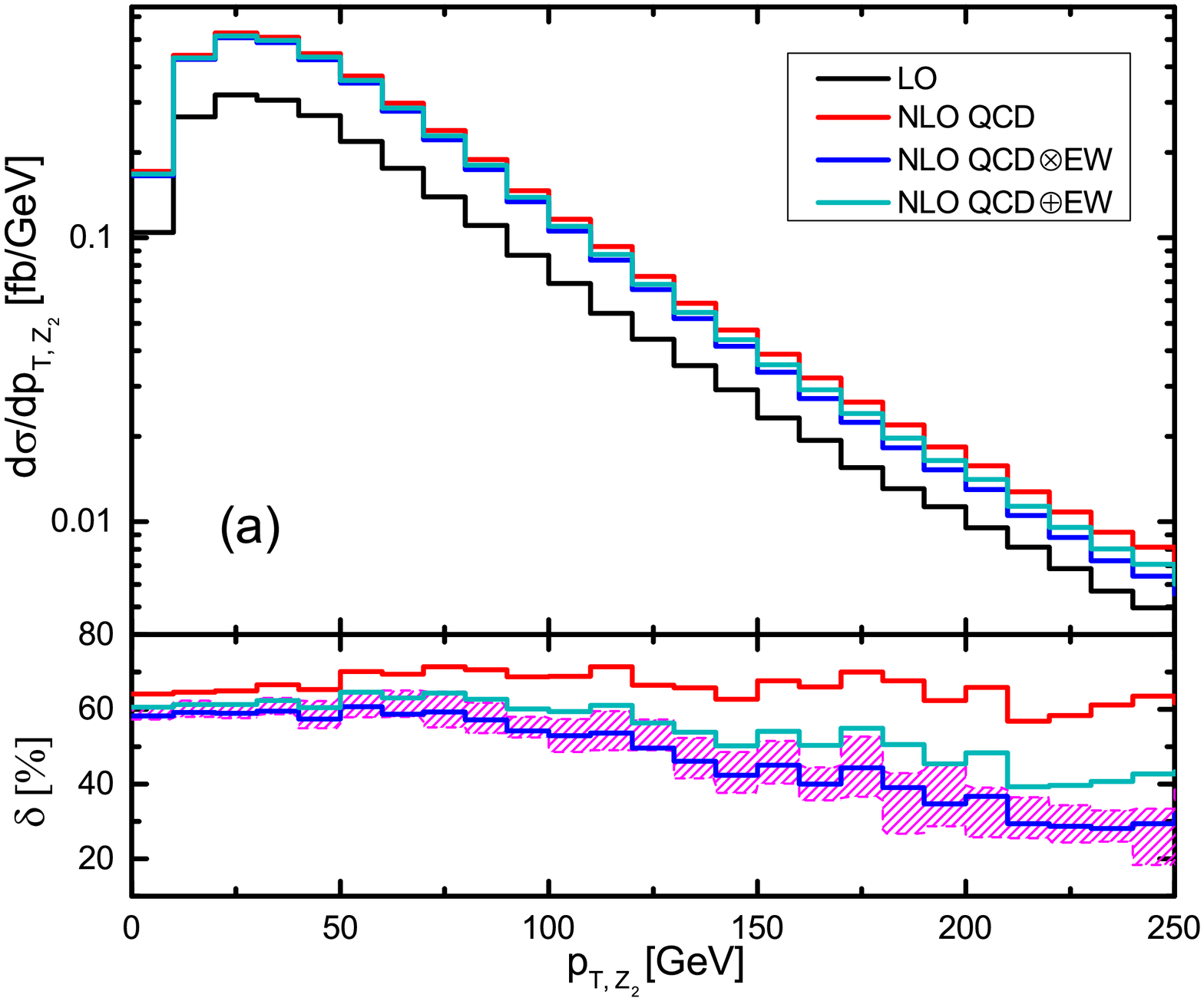}
\hspace{-1cm}
\includegraphics[width=9cm]{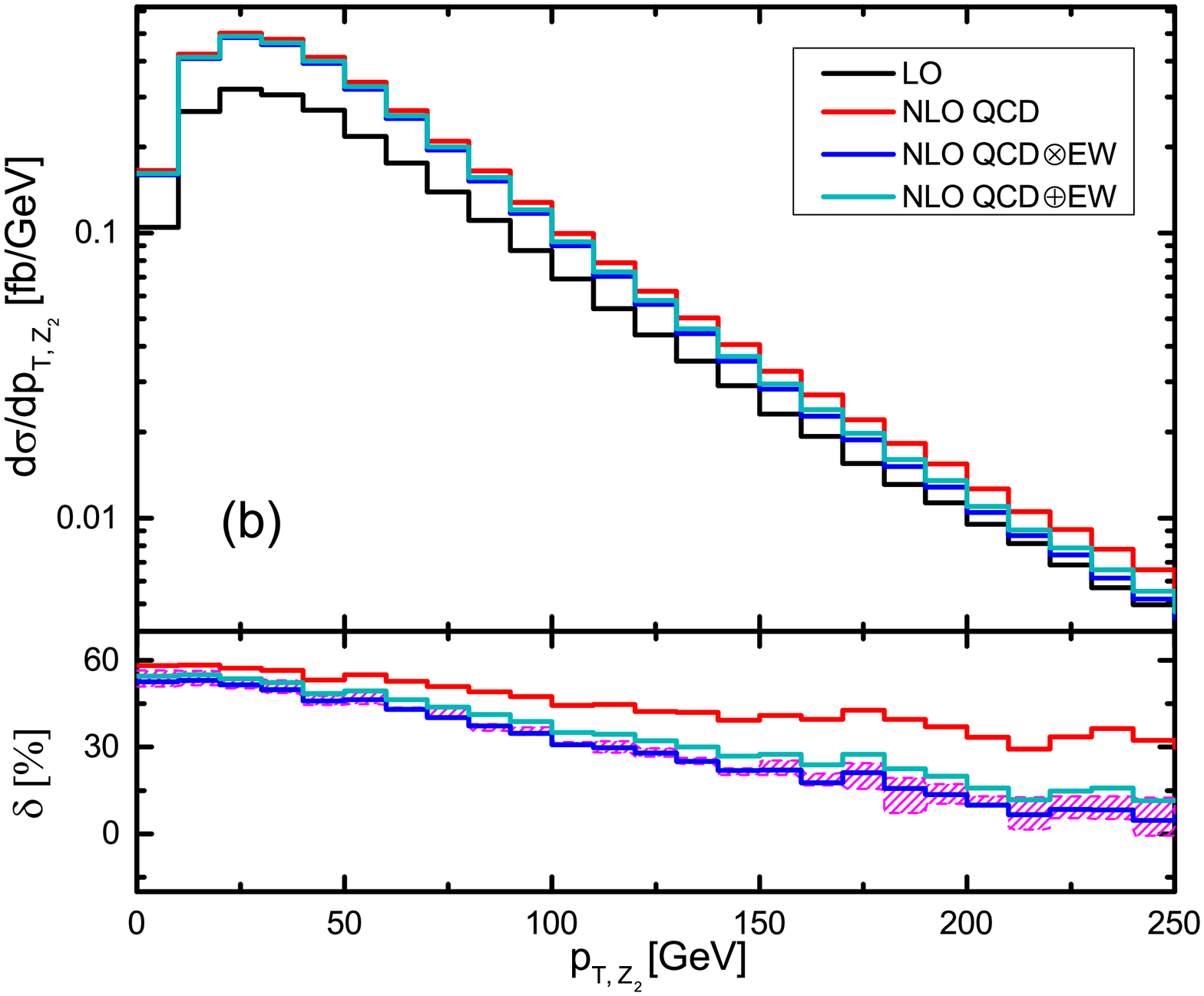}
\caption{ The LO, NLO QCD and NLO QCD+EW corrected distributions of the next-to-leading $Z$-boson transverse momentum for the inclusive (a) and exclusive (b) $ZZ\gamma$ production at the 14 TeV LHC. In the lower panels the pink bands are for the scale uncertainty ranges of the NLO QCD$\otimes$EW relative corrections, and the red, blue and green lines are for the NLO QCD, NLO QCD$\otimes$EW and NLO QCD$\oplus$EW relative correction distributions, respectively. }
\label{fig6}
\end{figure}

\par
We present the LO, NLO QCD and NLO QCD+EW corrected rapidity distributions of $Z$-boson pair in the inclusive and exclusive selection schemes in figure \ref{fig7}(a), (b), and the corresponding relative corrections are shown in the lower panels. From these figures we see that the NLO QCD relative correction in the inclusive scheme decreased from about $80\%$ to $50\%$ with the increment of $|y_{ZZ}|$, while the EW correction reduces obviously the related QCD corrected rapidity distributions in both event selection schemes in the whole plotted range. We find that the EW relative correction in the inclusive scheme is, however, insensitive to $|y_{ZZ}|$, and the inclusive NLO QCD$\otimes$EW (NLO QCD$\oplus$EW) relative correction varies from about $65\%~(70\%)$ to $40\%~(45\%)$ as $|y_{ZZ}|$ goes from $0$ to $2.5$. The NLO QCD relative correction in the exclusive scheme varies from about $60\%$ to $40\%$ in the plotted region, and consequently we obtain its full NLO QCD$\otimes$EW relative correction with the value varying in the range of $[30\%,~50\%]$ and the difference between the NLO QCD$\otimes$EW and NLO QCD$\oplus$EW never exceed 5\%.
\begin{figure}[htbp]
\includegraphics[width=9cm]{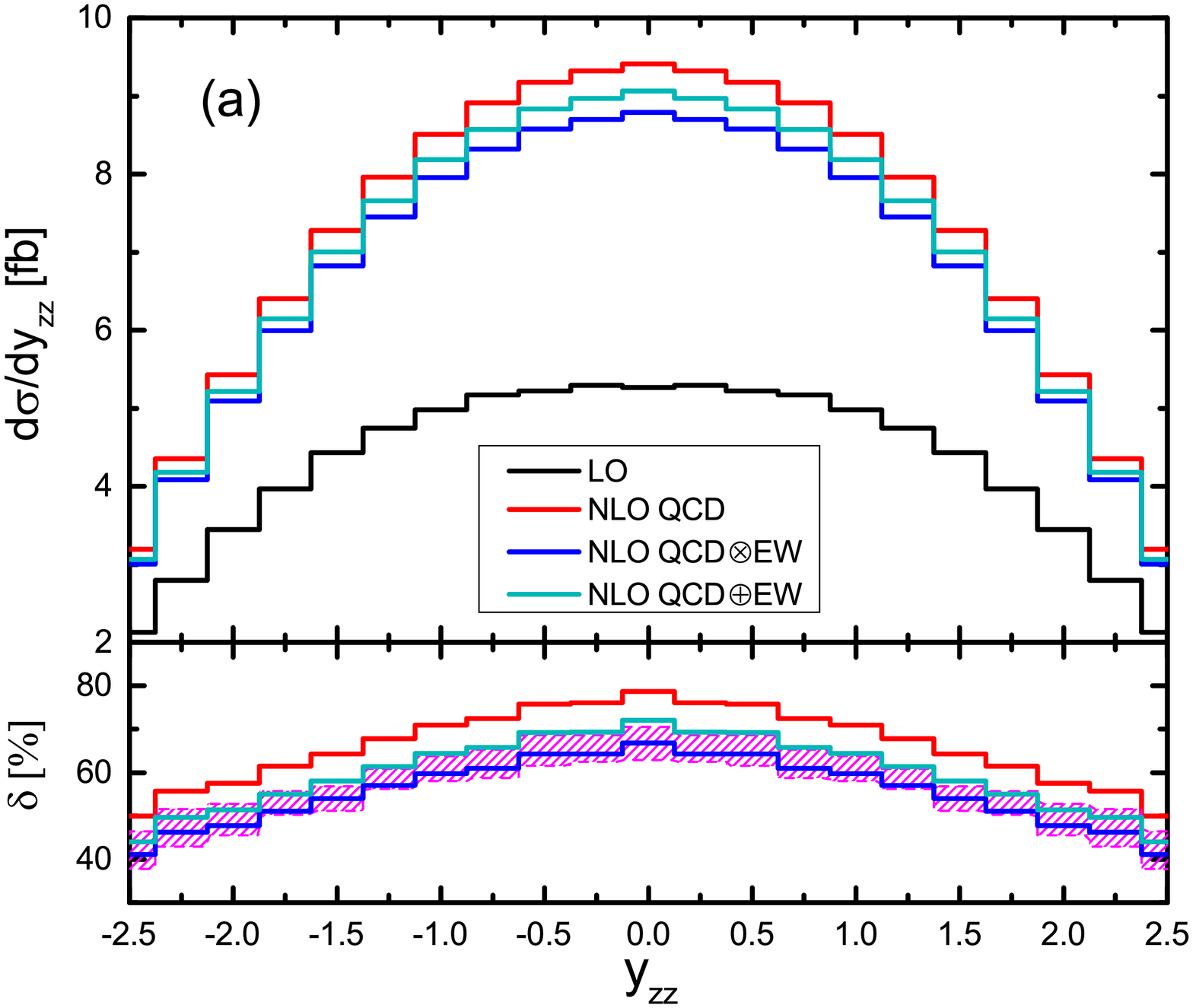}
\hspace{-1cm}
\includegraphics[width=9cm]{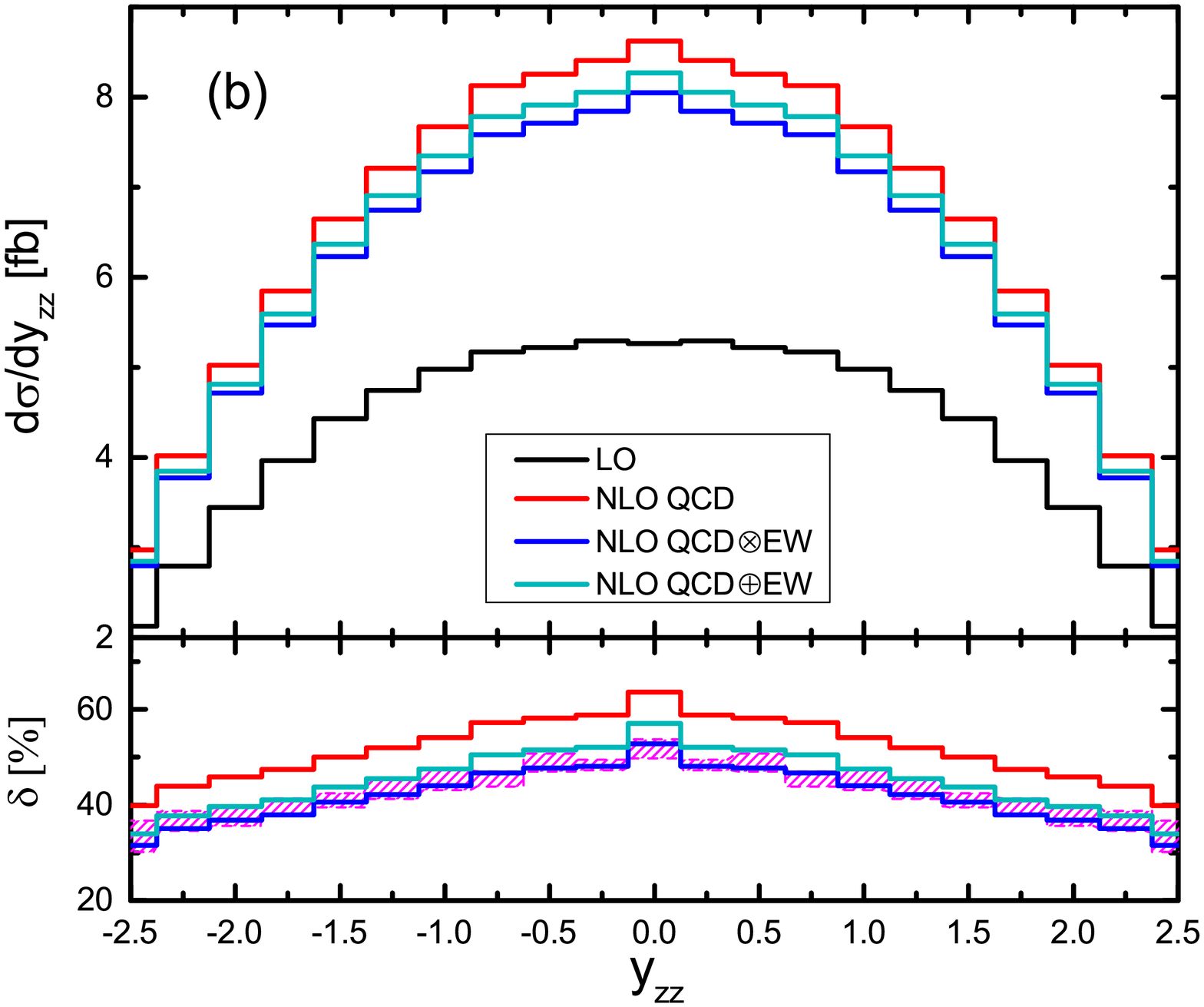}
\caption{ The $Z$-boson pair rapidity distributions at the LO, including NLO QCD and NLO QCD+EW corrections for the inclusive (a) and exclusive (b) $ZZ\gamma$ production at the 14 TeV LHC. In the lower panels the pink bands are for the scale uncertainty ranges of the NLO QCD$\otimes$EW relative corrections, and the red, blue and green lines are for the NLO QCD, NLO QCD$\otimes$EW and NLO QCD$\oplus$EW relative correction distributions, respectively. }
\label{fig7}
\end{figure}

\par
\subsubsection{Z-boson leptonic decays in \sc{MadSpin}}
\par
Normally, we can employ the NWA to calculate the $ZZ\gamma$ production with the subsequent $Z$-boson leptonic decays, i.e., $pp \to ZZ\gamma + X \to \ell_1^{+} \ell_1^{-} \ell_2^{+} \ell_2^{-} \gamma + X$, $(\ell_1,\ell_2= e, \mu, \tau)$. The so-called naive-NWA method simply produces final lepton pairs isotropically in the on-shell $Z$-boson central-of-mass system. However, the information of the non-trivial angular correlation due to the spin-correlation and $Z$-boson off-shell effects is sacrificed in this method. With an improved accuracy and also acceptable efficiency, an improved NWA method called {\sc MadSpin} \cite{Madspin} is developed, which is implemented as a generic tool in the {\sc MadGraph5} package based on the Frixione-Laenen-Motylinski-Webber (FLMW) method \cite{FLMW}. In fact, the {\sc MadSpin} can only retain spin-correlation effects at tree-level accuracy, the loop amplitude information can be empirically ignored and thus replaced by the corresponding tree-level amplitude (athough systematic further research may needed in the NLO EW calculation). In the following we discuss the approximate effects of the NLO QCD and NLO QCD$\otimes$EW corrections to $pp \to ZZ\gamma + X \to \ell_1^{+} \ell_1^{-} \ell_2^{+} \ell_2^{-} \gamma + X$ at the 14 TeV LHC, where the $Z$-boson leptonic decays are dealt with only at LO using {\sc MadSpin}. However, the NLO EW corrections to the $Z$-boson leptonic decays, in particular the final-state radiation (FSR) of photons from charged leptons, may induce sizable effects. Based on the results and discussions on $W/Z+\gamma$ production at the LHC in \cite{w+gamma,z+gamma}, it is expected that the FSR contributions are considerable, and are even dominant over the NLO EW corrections from production only.

\par
We name the leading negative charged lepton $\ell_1^-$ and the next-to-leading negative charged lepton $\ell_2^-$ according to their transverse momentum magnitudes, i.e, $p_{T,\ell_1^-} > p_{T,\ell_2^-}$. In figures \ref{fig8}(a), (b) and figures \ref{fig9}(a), (b) we depict the leading/next-to-leading negative charged lepton transverse momentum distributions at the LO, including the NLO QCD and NLO QCD$\otimes$EW corrections to the $ZZ\gamma$ production separately, and the relative corrections are shown in the lower panels. Clearly the inclusion of EW correction modifies the LO and NLO QCD corrected distributions in both the inclusive and exclusive schemes. For the leading negative charged lepton, the NLO QCD relative correction in the inclusive scheme ascends in the range of $[60\%,90\%]$ with the $p_{T, \ell_1^-}$ $\in$ $[20,180] \rm ~GeV$, while in the exclusive scheme the decreased tendency is shown from about $55\%$ to $40\%$. The inclusion of the NLO EW correction makes the NLO QCD$\otimes$EW relative correction in the inclusive scheme rather flat under $60\%$, and the NLO QCD$\otimes$EW relative correction in the exclusive scheme varies from $50\%$ to $20\%$ in the plotted region. Similarly the next-to-leading negative charged lepton receives a large negative EW correction in high $p_{T, \ell_2^-}$ region, numerically the NLO QCD$\otimes$EW relative correction in inclusive/exclusive scheme is $60\%$/$45\%$ at $p_{T, \ell_2^-}\sim60~{\rm GeV}$ and $30\%$/$10\%$ at $p_{T, \ell_2^-}\sim 140~{\rm GeV}$.
\begin{figure}[htbp]
\includegraphics[width=9cm]{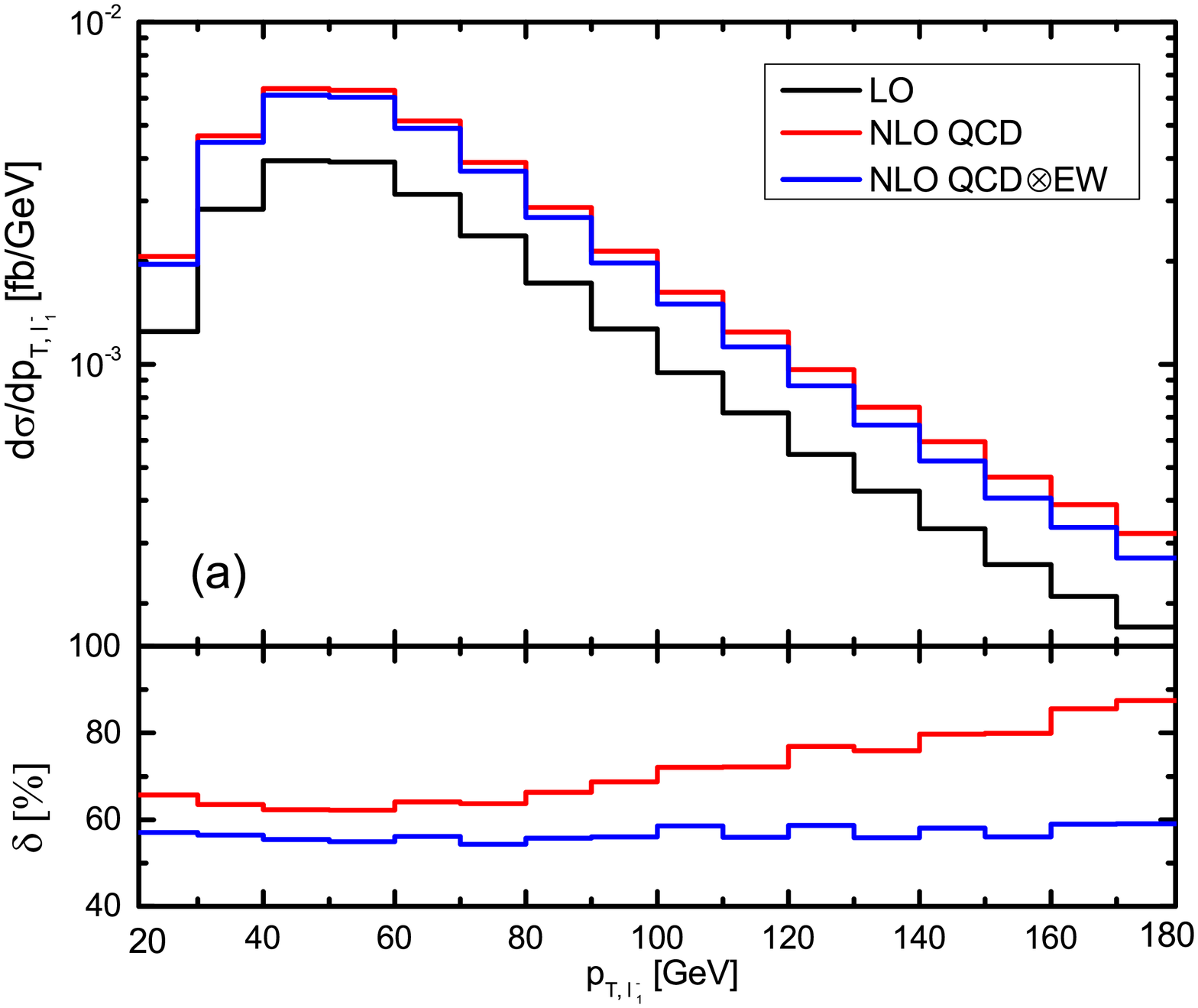}
\hspace{-1cm}
\includegraphics[width=9cm]{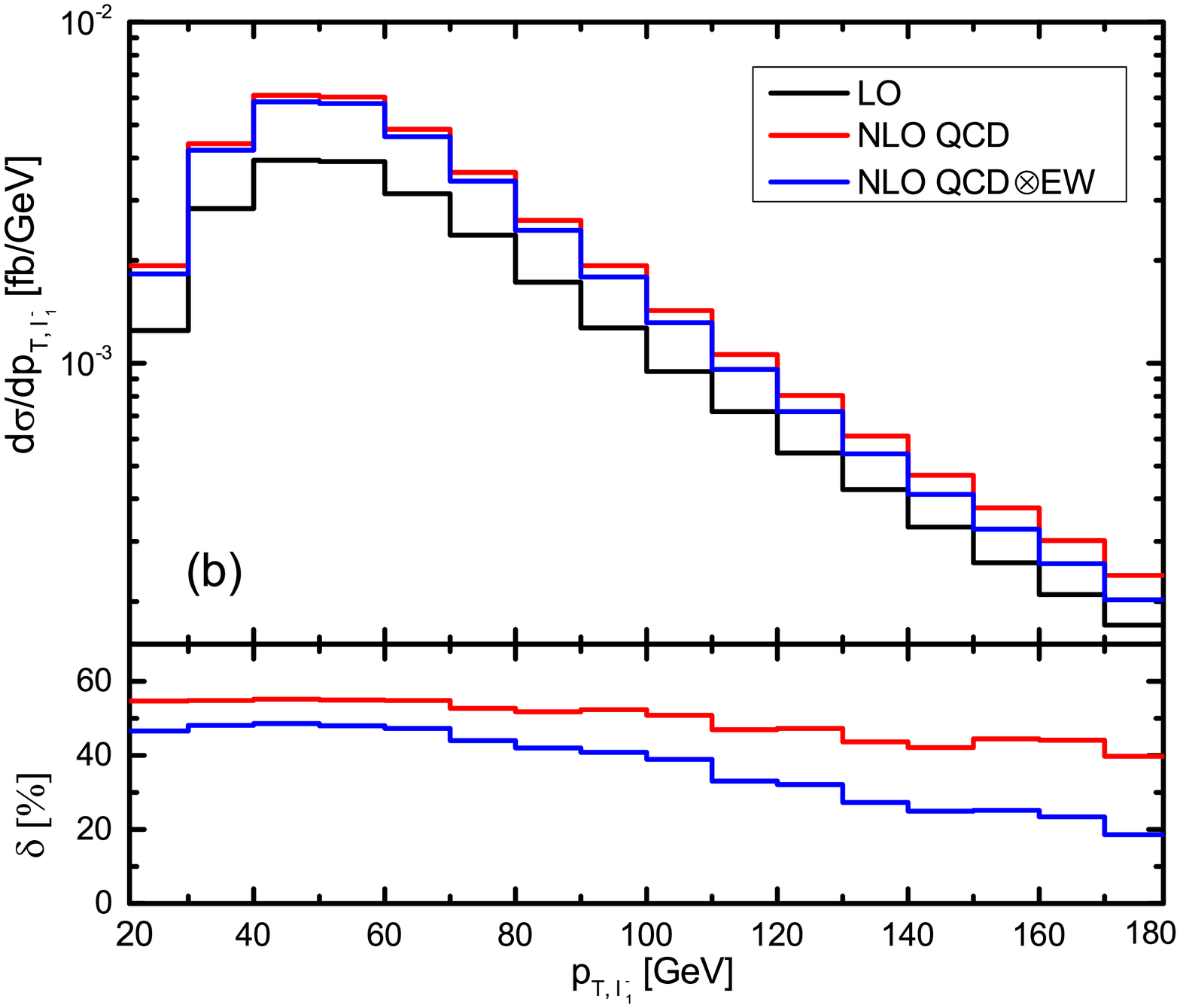}
\caption{ The leading negative charged lepton tranverse momentum distributions at the LO, including NLO QCD and NLO QCD$\otimes$EW combined corrections to $pp \to ZZ\gamma \to \ell_1^{+} \ell_1^{-} \ell_2^{+} \ell_2^{-} \gamma + X$ $(\ell_1, \ell_2= e, \mu, \tau)$ in the inclusive (a) and exlucsive (b) schemes at the 14 TeV LHC. In the lowers panels the red lines are for the NLO QCD relative correction distributions, and the blue lines are for the NLO QCD$\otimes$EW combined relative correction distributions.}
\label{fig8}
\end{figure}
\begin{figure}[htbp]
\includegraphics[width=9cm]{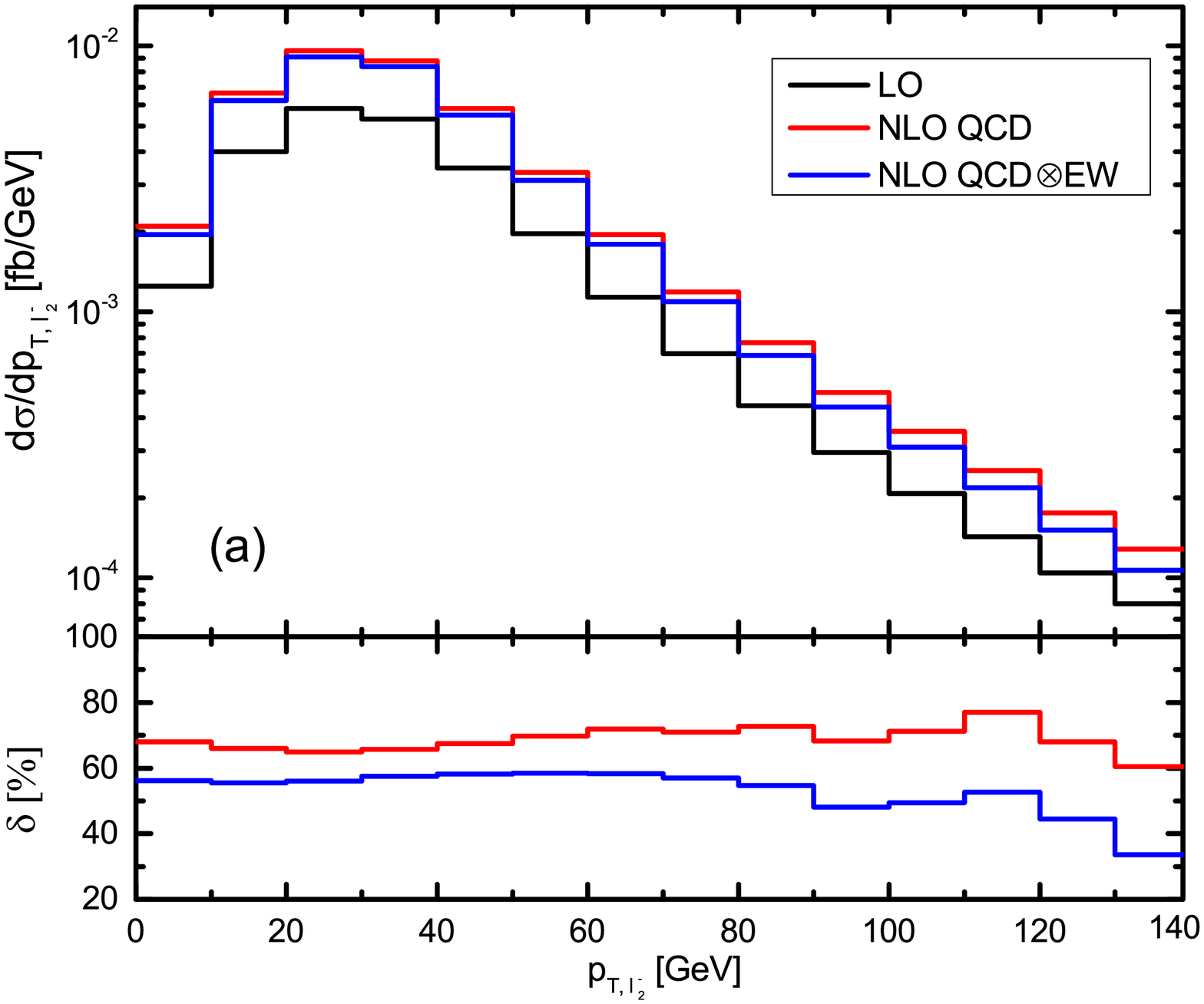}
\hspace{-1cm}
\includegraphics[width=9cm]{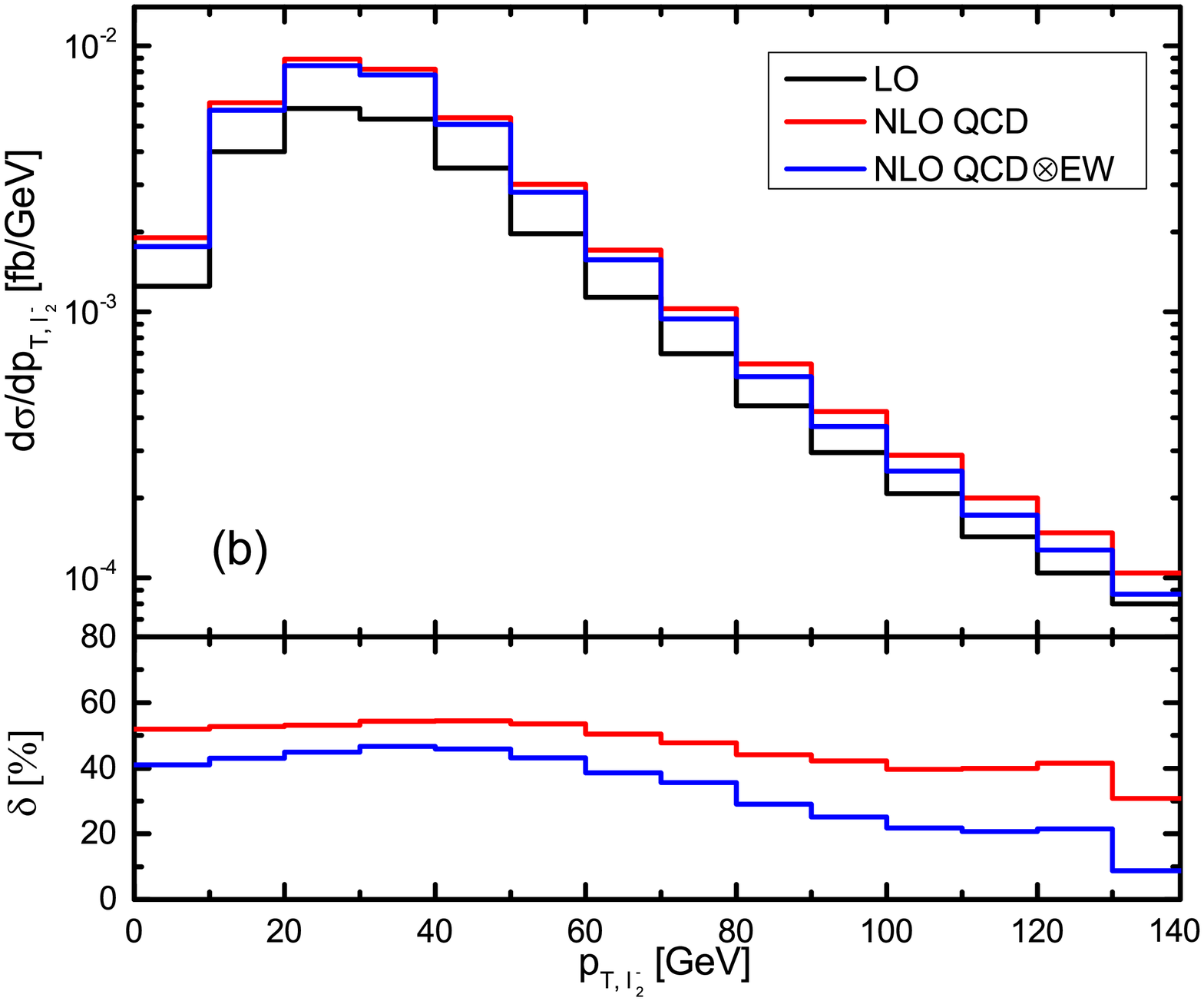}
\caption{ The next-to-leading negative charged lepton tranverse momentum distributions at the LO, including NLO QCD and NLO QCD$\otimes$EW combined corrections to $pp \to ZZ\gamma \to \ell_1^{+} \ell_1^{-} \ell_2^{+} \ell_2^{-} \gamma + X$ $(\ell_1, \ell_2= e, \mu, \tau)$ in the inclusive (a) and exlucsive (b) schemes at the 14 TeV LHC. In the lowers panels the red lines are for the NLO QCD relative correction distributions, and the blue lines are for the NLO QCD$\otimes$EW combined relative correction distributions.}
\label{fig9}
\end{figure}
\par
In figure \ref{fig10}(a), (b) we present the LO, NLO QCD and NLO QCD$\otimes$EW corrected distributions of the azimuthal-angle difference between the final two negative charged leptons ($\phi_{\ell_1^- \ell_2^-}$), and the corresponding relative corrections, respectively. These figures demonstrate that the EW relative correction to the $ZZ\gamma$ production has less relevance to $\phi_{\ell_1^- \ell_2^-}$. In the inclusive scheme the NLO QCD relative correction is stable at about $65\%$ and the NLO QCD$\otimes$EW relative correction is reduced to about $60\%$ in the plotted range, while in the exclusive scheme the NLO QCD/NLO QCD$\otimes$EW relative correction is slightly increased compared to the inclusive scheme, and has the value of about $50\%$/$40\%$ at $\phi_{\ell_1^- \ell_2^-}=0$ and 60\%/40\% at $\phi_{\ell_1^- \ell_2^-}=\pi$.
\begin{figure}[htbp]
\includegraphics[width=9cm]{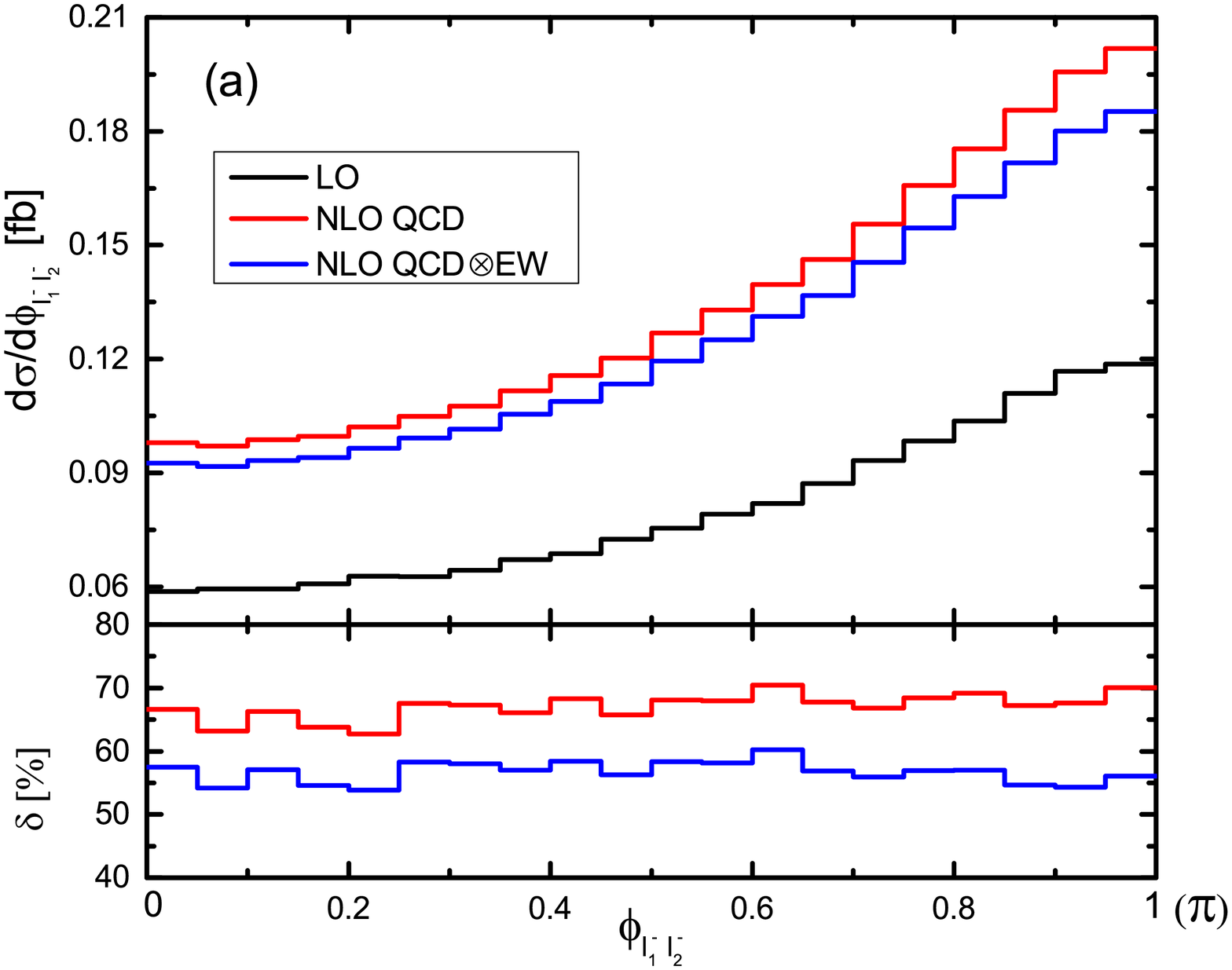}
\hspace{-1cm}
\includegraphics[width=9cm]{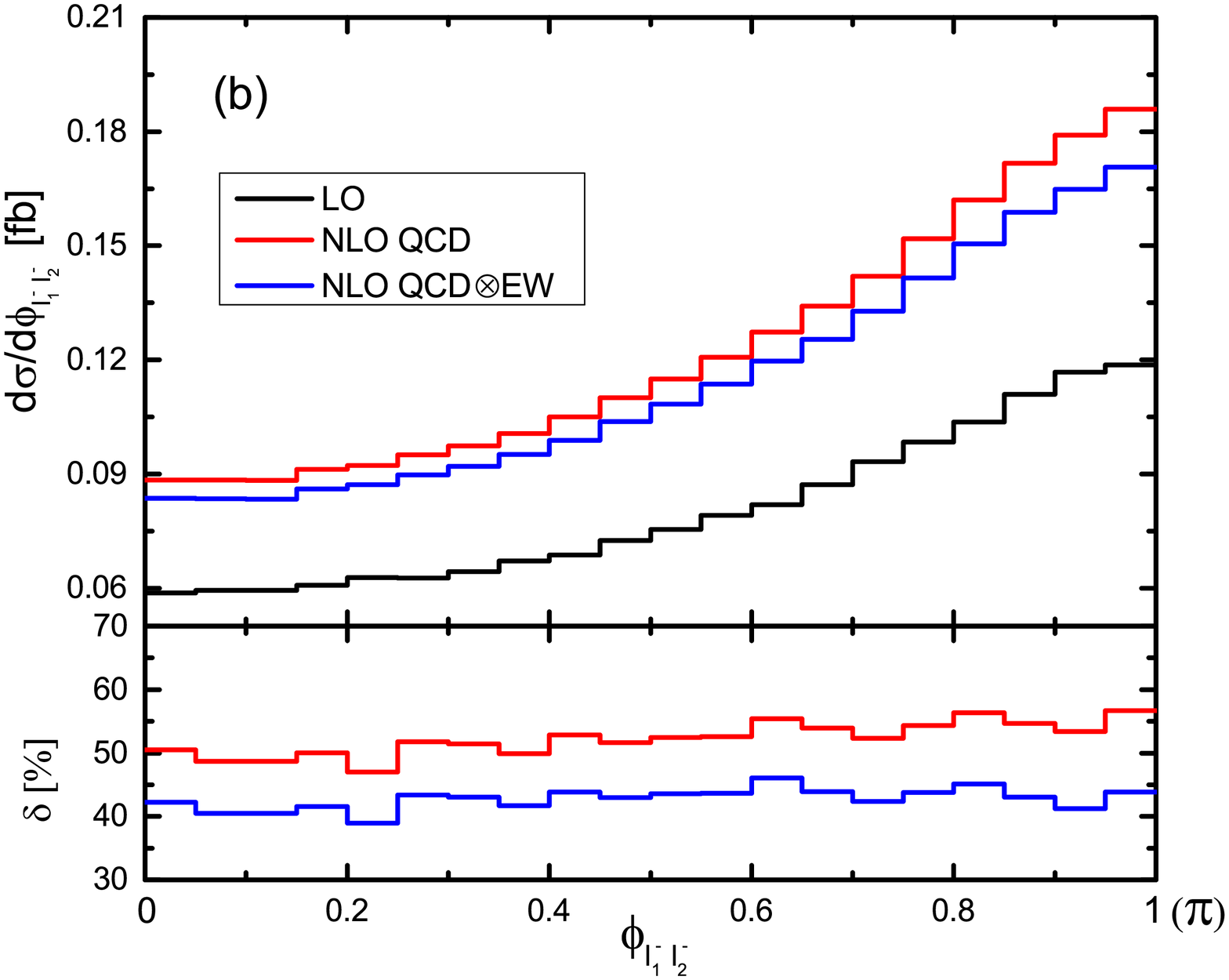}
\caption{ The azimuthal-angle difference distributions between the two negative charged leptons at the LO, including NLO QCD and NLO QCD$\otimes$EW combined corrections to $pp \to ZZ\gamma \to \ell_1^{+} \ell_1^{-} \ell_2^{+} \ell_2^{-} \gamma + X$ $(\ell_1, \ell_2= e, \mu, \tau)$ in the inclusive (a) and exlucsive (b) schemes at the 14 TeV LHC. In the lowers panels the red lines are for the NLO QCD relative correction distributions, and the blue lines are for the NLO QCD$\otimes$EW combined relative correction distributions.}
\label{fig10}
\end{figure}

\vskip 5mm
\section{Summary}  \label{summary}
\label{summary}
\par
In this work we study for the first time the impact of the NLO EW correction to the $pp \to ZZ \gamma+X$ process at the LHC. The subsequent $Z$-boson leptonic decays are disposed by using {\sc MadSpin} method considering the spin-correlation and off-shell effects at LO accuracy. Various integrated cross sections and kinematic distributions are provided at the NLO QCD+EW accuracy. We find that although the NLO EW correction to a certain physical observable of this production is smaller than the relevant NLO QCD correction, it is still significant and needs to be taken into account in future high-precision  experimental analysis. In our calculation we perform the separation of a final-state photon-jet system in real emission processes by using both the Frixione isolated-photon plus jets algorithm and the phenomenological quark-to-photon fragmentation function method for comparison. Our results show that the scale dependence can be slightly reduced by the inclusion of NLO QCD+EW correction, and can further be weakened by applying the jet-veto selection scheme. However, the jet-veto would induce an additional theoretical uncertainty which can be improved by adopting the resummation technique.

\vskip 5mm
\section*{Acknowledgments}
This work was supported in part by the National Natural Science Foundation of China (No.11375171, No.11405173, No.11535002).





\vskip 5mm

\begin{thebibliography}{99}
\bibitem{Higgs1}
Aad G \emph{et al} (ATLAS Collaboration) 2012 \emph{Phys. Lett.} B 716 1

\bibitem{Higgs2}
Chatrchyan S \emph{et al} (CMS Collaboration) 2012 \emph{Phys. Lett.} B 716 30
\bibitem{ATLAS}
Aad G \emph{et al} (ATLAS Collaboration) 2014 \emph{Phys. Lett.} B 732 8

\bibitem{CMS}
Chatrchyan S \emph{et al} (CMS Collaboration) 2013 \emph{Phys. Lett.} B 726 587
\bibitem{ZZr-NLO-QCD}
Bozzi G, Campanario F, Hankele F and Zeppenfeld D 2013 \emph{Phys. Rev.} D 81 094030

\bibitem{wishlt2013}
Butterworth J \emph{et al} [arXiv:hep-ph/1405.1067]

\bibitem{wishlt2015}
Badger S \emph{et al} [arXiv:hep-ph/1605.04692]

\bibitem{Feynarts}
Hahn T 2001 \emph{Comput. Phys. Commun.} 140 418

\bibitem{Formcalc}
Hahn T Perez-Victoria M 1999 \emph{Comput. Phys. Commun.} 118 153

\bibitem{Looptools}
van Oldenborgh G J 1991 \emph{Comput. Phys. Commun.} 66 1

\bibitem{dipole}
Catani S and Seymour M H 1997 \emph{Nucl. Phys. B} 485 291

\bibitem{tcpss}
Harris B W and Owens J F 2002 \emph{Phys. Rev.} D  65 094032

\bibitem{Madgraph5}
Alwall J \emph{et al} 2014 \emph{J. High Energy Phys.} JHEP07(2014)079

\bibitem{w+gamma}
Denner A, Dittmaier S, Hecht M and Pasold C 2015  \emph{J. High Energy Phys.} JHEP04(2015)018

\bibitem{z+gamma}
Denner A, Dittmaier S, Hecht M and Pasold C 2016  \emph{J. High Energy Phys.} JHEP02(2016)057

\bibitem{Sudakov}
Denner A and Pozzorini S 2001 \emph{Eur. Phys. J.} C 18 461
Denner A and Pozzorini S 2001 \emph{Eur. Phys. J.} C 21 63

\bibitem{alpha0-input}
Denner A 1993 \emph{Fortschr. Phys.} 41 307

\bibitem{delta}
Sirlin A 1980 \emph{Phys. Rev.} D 22 971

\bibitem{NNPDF}
Ball R D \emph{et al} (NNPDF Collaboration) 2013 \emph{Nucl. Phys.} B 877 290

\bibitem{Denner}
Denner A and Dittmaier S 2003 \emph{Nucl. Phys.} B 658 175

\bibitem{PV}
Passarino G and Veltman M 1997 \emph{Nucl. Phys.} B 160 151

\bibitem{detG1}
Nhung D T, Ninh L D, and Weber M M 2013 \emph{J. High Energy Phys.} JHEP12(2013)096

\bibitem{detG2}
Su J-J, Ma W-G, Zhang R-Y, Wang S-M, and Guo L 2008 \emph{Phys. Rev.} D 78 016007

\bibitem{zzj}
Wang Y, Zhang R-Y, Ma W-G, Li X-Z, Guo L 2016 \emph{Phys. Rev.} D 94 013011

\bibitem{LEP1}
Glover E W N and Morgan A G 1994 \emph{Z. Phys.} C 62 311

\bibitem{LEP2}
Buskulic D \emph{et al} (ALEPH Collaboration) 1996 \emph{Z. Phys.} C 69 365

\bibitem{Frixione}
Frixione S 1998 \emph{Phys. Lett.} B 429 369

\bibitem{PDG}
Patrignani C \emph{et al} (Particle Data Group) 2016 \emph{Chin. Phys.} C 40 100001

\bibitem{DIS-nsct}
Diener K-P O, Dittmaier S, Hollik W 2005 \emph{Phys. Rev.} D 72 093002

\bibitem{jt-uncertainty}
Stewart I W and Tackmann F J 2012 \emph{Phys. Rev.} D 85 034011

\bibitem{LUXqed}
Manohar A, Nason P, Salam G P, and Zanderighi G 2016 \emph{Phys. Rev. Lett.} 117 242002

\bibitem{Madspin}
Artoisenet P, Frederix R, Mattelaer O, and Rietkerk R 2013 \emph{J. High Energy Phys.} JHEP03(2013)015

\bibitem{FLMW}
Frixione S, Laenen E, Motylinski P, and Webber B R 2007 \emph{J. High Energy Phys.} JHEP04(2007)081


\end{thebibliography}
\end{document}